\documentclass[12pt]{article}

\usepackage{amsmath}
\usepackage{amssymb}
\usepackage{latexsym}
\usepackage{graphicx}
\usepackage{psfrag,fancyhdr,epsfig}

\def\beq{\begin{equation}}
\def\eeq{\end{equation}}
\def\bea{\begin{eqnarray}}
\def\eea{\end{eqnarray}}
\begin{document}

\begin{titlepage}

\vspace*{1cm}
\begin{center}
{\bf \Large {Angular profile of Particle Emission
from a\\[2mm] Higher-dimensional Black Hole: Analytic Results}}

\bigskip \bigskip \medskip

{\bf Panagiota Kanti} and {\bf Nikolaos Pappas}

\bigskip
{\it Division of Theoretical Physics, Department of Physics,\\
University of Ioannina, Ioannina GR-451 10,  \\ Greece}

\begin{abstract}
During the spin-down phase of the life of a higher-dimensional black hole, the
emission of particles on the brane exhibits a strong angular variation with
respect to the rotation axis of the black hole. It has been suggested that
this angular variation is the observable that could disentangle the dependence
of the radiation spectra on the number of extra dimensions and angular momentum
of the black hole. Working in the low-energy regime, we have employed analytical
formulae for the greybody factors, angular eigenvalues and eigenfunctions of
fermions and gauge bosons, and studied the characteristics of the corresponding
angular profiles of emission spectra in terms of only a few dominant partial
modes. We have confirmed that, in the low-energy channel, the emitted gauge
bosons become aligned to the rotation axis of the produced black hole while
fermions form an angle with the rotation axis whose exact value depends on
the angular-momentum of the black hole. In the case of scalar fields, we
demonstrated the existence of a ``spherically-symmetric zone" that is followed
by the concentration of the emission on the equatorial plane, again in total
agreement with the exact numerical results.
\end{abstract}

\end{center}

\end{titlepage}




\section{Introduction}

Under the assumption that a low-energy scale for gravity exists in the
context of a higher-dimensional fundamental theory \cite{ADD}, the possibility
of observing in the near future quantum-gravity effects has excited a
lot of interest among high-energy physicists, both theorists and experimentalists.
The main reason for that is the fact that, if $M_*$ -- the fundamental gravity
scale -- is as low as a few TeV, then these effects could be observed during
trans-Planckian particle collisions at current, ground-based accelerators \cite{BF}.
One such strong-gravity effect could be the creation of higher-dimensional
miniature black holes during the collision of ordinary Standard-Model particles
localised on our brane -- a (3+1)-dimensional hypersurface embedded in the 
$(4+n)$-dimensional spacetime, the bulk. Due to their small size, these
black holes will have a high temperature and will evaporate very quickly
via Hawking radiation \cite{hawking}, i.e. the emission of ordinary particles
with a thermal spectrum \cite{Kanti, reviews, Harris}. 

The emission of Hawking radiation is anticipated to take place during the two
intermediate phases in the life of the black hole, the spin-down and the 
Schwarzschild phase. It is expected to be the main observable signal not
only of the creation of these miniature black holes but of the existence
of the extra spacelike dimensions themselves in the absence of which the
creation of the former would not be possible. As a result, the study of the
emission of Hawking radiation by higher-dimensional black holes has been
intense during the last ten years. In the early days, the Schwarzschild
phase -- the spherically-symmetric phase in the life of the black hole
arising presumably after the shedding of its angular momentum -- was
considered to be the longest and thus the most important. It was also
the one with the simplest metric tensor describing the spacetime around
it, and therefore the first one to be exhaustively studied both analytically
\cite{kmr1, Frolov1} and numerically \cite{HK1, graviton-schw}. The
results derived showed a strong dependence of the emission rates of
all types of Standard Model particles on the brane on the number of
spacelike dimensions existing transversely to the
brane~\footnote{Variants of the spherically-symmetric Schwarzschild phase,
where a cosmological constant \cite{KGB} or the
higher-curvature Gauss-Bonnet term \cite{GBK} were introduced, were also 
studied with the spectrum exhibiting a dependence also on parameters
related to these terms. In addition, the Schwarzschild phase of
quantum-corrected black holes has been studied in \cite{Nicolini}.}.

One was thus led to hope that by detecting the emission of Hawking radiation
could not only shed light on aspects arising from the interplay between
classical gravity and quantum physics but also give a quantitative answer
to a century-old fundamental question, that of the dimensionality of spacetime. 
Nevertheless, the Schwarzschild phase is preceded by the axially-symmetric
spin-down phase. The gravitational background around a simply-rotating
black hole -- one of the very few cases where the equations of motion
of the propagating particles can be decoupled and solved -- depends also
on the angular-momentum parameter $a$ of the black hole. According to the
results existing in the literature \cite{DHKW, CKW, CDKW1, IOP, CEKT2, CEKT3,
rot-other, CEKT4, CDKW2, graviton-rot, brane-bulk, Sampaio, KP1}, this
dependence is carried over in
the form of the radiation emission spectra and is, in fact, found to be
similar to the effect that the number of additional spacelike dimensions $n$
has on them. To complicate things more, simulations of black hole events
\cite{charybdis2, blackmax} have revealed that the spin-down phase is not
a short-lived one, as previously thought, and that the rotation of the
black hole remains significant for most of its lifetime.

The fact that the dependence of the radiation spectra on the number of extra
dimensions $n$ for all types of particles is entangled with the dependence
on the angular-momentum parameter $a$ means that measuring both of these
parameters is extremely difficult. The only way out was to employ another
observable that would strongly depend on only one of these two parameters 
while being insensitive to the other. Upon determination of that particular
parameter, the second could then be determined from the radiation spectra. One
characteristic feature of the emission spectra coming from the spin-down
phase is the non-isotropic emission, in contrast to the one coming from
the Schwarzschild phase where the emitted particles are evenly distributed
over a $4\pi$ solid angle. It has therefore been suggested \cite{FST, CDKW3}
that this non-isotropy can serve as the additional observable necessary to
disentangle the $n$ and $a$-dependence of the spectra. Indeed, it was
demonstrated \cite{CDKW3} that the angular profile of the emitted radiation
depends extremely weakly on the number of additional dimensions $n$ while
it may provide valuable information on the angular momentum of the black hole
(see, for example, \cite{Sampaio-ang}).

More specifically, under the combined effect of the centrifugal force exerted
on the emitted particles and the spin-rotation coupling for particles with
non-zero spin (an analytical explanation of the latter is given in
\cite{Stojkovic-ang}), the orientation of the emitted radiation depends strongly on
the energy channel in which the particles are emitted and on how fast the black
hole rotates. If we look specifically at the low-energy channel, then
we observe that gauge bosons and fermions have a distinctly different behaviour:
the emitted gauge bosons remain aligned to the rotation axis of the black hole
independently of the angular-momentum parameter; fermions, on the other hand,
form an angle with the rotation axis whose value strongly depends on the
value of $a$. As a result, the orientation of gauge bosons can serve as a
good indicator of the rotation axis of the black hole \cite{CDKW3} and the
orientation of fermions can then provide a measurement of the value of the
angular momentum of the black hole \cite{FST, CDKW3}.

The aforementioned results presented in \cite{FST, CDKW3} were derived by means
of a very complicated and time-consuming process that involved the numerical 
integration of both the radial and angular part of the equation of motion of
each emitted particle as well as additional challenges such as the numerical
calculation of the angular eigenvalue itself, which does not exist in closed
form for a rotating background, and the summation of a very large number of
partial modes. The purpose of this work is to provide an alternative way of
deriving the angular profile of the emitted radiation without resorting to
complicated numerical calculations. This is facilitated by the fact that all
valuable information that may be derived from the angular spectra is restricted
in the low-energy regime where the radial equations for all types of particles
have been analytically solved \cite{CEKT2, CEKT3}. In addition, analytical
formulae, in the form of power series, for the angular eigenfunction and
eigenvalue exist in the literature. By combining all the above in a
constructive way, we investigate which contributions are the dominant ones,
that predominantly determine the angular profile of the emitted radiation.
In this way, we formulate simple constraints involving a finite number of terms
and partial modes that successfully reproduce all the features of the anisotropic
emission, namely the value of the angle where the emission becomes maximum and
the corresponding value of the energy emission rate.

The structure of this paper is as follows. In section \ref{sec-theory}, we 
present the theoretical framework with the field equations that need to be
solved and the corresponding energy emission rates for a general spin-$s$
field. In section \ref{sec-solutions}, we present the analytical formulae
for the greybody factors, angular eigenfunctions and eigenvalues that will
be our tools for the analytical investigation of the angular profile of
the emitted radiation. In section \ref{sec-profile}, we consider separately
the cases of fermions, gauge bosons and, for completeness, scalar fields too,
emitted by a higher-dimensional simply-rotating black hole on the brane:
in each case, we determine the dominant modes, formulate simple extremization
constraints with respect to the angle of emission $\theta$, and derive
their angular distribution on the brane.  Finally, in section \ref{sec-conclusions},
we summarise our results and present our conclusions.


\section{Theoretical framework \label{sec-theory}}

The most generic type of a black hole in a higher-dimensional spacetime is the
one that rotates around one or more axes. The gravitational field around 
such a black hole is described by the Myers-Perry solution \cite{MP}. However,
it is only for particular configurations of the angular-momentum components
that the equation of motion of a particle propagating in the higher-dimensional
spacetime can be decoupled into an angular and a radial part. The case of a
simply-rotating black hole, where the black hole possesses only one angular-momentum
component that lies on a plane parallel to our brane, corresponds to one of these
configurations and the one that has been mostly considered in the literature.
This choice is also justified by the assumption that the black hole, if 
created by the collision of two brane-localised particles, will acquire an angular
momentum component along the (3+1)-dimensional part of the full manifold. 

In this work, we will also focus on the case of a simply-rotating black hole. 
In addition, we will study effects that take place strictly on our brane, namely
the emission of Hawking radiation by the higher-dimensional, rotating black hole 
in the form of non-zero-spin Standard-Model fields. The line-element of the
brane background in which these particles propagate is given by the expression
\cite{Kanti}
\begin{eqnarray}
ds^2 & =& -\left( 1 - \frac{\mu}{\Sigma r^{n-1}} \right) dt^2
- \frac{2 a \mu \sin^2 \theta}
{\Sigma r^{n-1}}\,dt \, d\varphi
 + \left( r^2 + a^2 + \frac{a^2 \mu \sin^2 \theta}{\Sigma r^{n-1}} \right) \sin^2
\theta\,d \varphi^2
\nonumber \\ & &
+ \frac{\Sigma}{\Delta}\,dr^2
+ \Sigma\,d\theta^2
\,,
\label{metric}
\end{eqnarray}
where
\begin{equation}
\label{sigma}
\Delta = r^2 + a^2 - \frac{\mu}{r^{n-1}}, \quad \quad
\Sigma = r^2 + a^2 \cos^2 \theta\,.
\end{equation}
The mass  $M_{BH}$ of the black hole and its angular momentum $J$ are 
related to the $\mu$ and $a$ parameters, respectively, through the
relations
\begin{equation}
M_{BH} = \frac{(n+2) A_{n+2}}{16 \pi G}\,\mu , \quad \quad
J = \frac{2}{n+2} M_{BH}\,a\,,
\end{equation}
where $A_{n+2} = 2\pi^{(n+3)/2} / \Gamma[(n+3)/2]$ is the area of
an $(n+2)$-dimensional unit sphere, and $G$ is the $(4+n)$-dimensional
Newton's constant. The horizon radius $r_h$ follows from the equation
$\Delta(r) = 0$: for $n \ge  1$, it may be shown that there is only one real,
positive root, which may be implicitly written as $r_h^{n+1}=\mu/(1+a_*^2)$,
where $a_*$ is defined as $a_* \equiv a/r_h$.

The derivation of the field equations that the brane-localized Standard-Model
fields satisfy in the above background follows the analysis performed originally
by Teukolsky in 4 dimensions \cite{Teukolsky}. The method demands the use of the
Newman-Penrose formalism and results in a `master' partial differential equation
that scalars, fermions and gauge bosons obey on the brane. If we use a
factorized ansatz for the field perturbation $\Psi_h$ of the form
\begin{equation}
\Psi_h(t,r,\theta, \varphi)  =
\sum _{\Lambda }
{}_h a_\Lambda \,_hR_\Lambda(r)\,_hS_\Lambda(\theta)\,e^{-i\omega t}\,
e^{im\varphi}\,,
\label{fact}
\end{equation}
the aforementioned `master' equation separates, in the background of Eq. (\ref{metric}),
into two decoupled ordinary differential equations, a radial
\begin{equation}
\Delta^{-h}\frac{d\,}{dr}\left(\Delta^{h+1} \frac{d_hR_\Lambda}{dr}\right) +
\left[\frac{K^2-ihK \Delta'(r)}{\Delta}
+4ih\omega r + h(\Delta''(r)-2) \delta_{h,|h|} -
{}_h\lambda_\Lambda\right] {}_hR_\Lambda=0\,, \label{radial}
\end{equation}
and an angular one
\begin{equation}
\frac{d\,}{dx}\left[(1-x^2) \frac{d {}_hS_\Lambda(x)}{dx}\right]
+\left[a^2 \omega^2 x^2-2ha\omega x-
\frac{(m+hx)^2}{1-x^2} + h + {}_hA_\Lambda\right] {}_hS_\Lambda(x)=0\,.
\label{angular}
\end{equation}
In the above, $h$ is the spin-weight, $h=(-|s|,+|s|)$, of the given field that 
distinguishes its radiative components,  and $\Lambda=\{lm\omega\}$
denotes the set of `quantum numbers' of each mode. We have also defined
the quantities $K \equiv (r^2+a^2) \omega -am$ and $x \equiv\cos \theta$. Finally, 
${}_hA_\Lambda$ is the eigenvalue of the spin-weighted spheroidal harmonics
${}_hS_\Lambda(x)$ - as we will shortly comment, the value of this constant does
not exist in closed form. This quantity also determines the separation constant
between the radial and angular equations with 
${}_h\lambda_\Lambda \equiv {}_hA_\Lambda - 2ma\omega + a^2 \omega^2$.

The above set of equations has been used in the literature in order to study
the emission of Hawking radiation, in the form of an arbitrary spin-$s$ field,
from a higher-dimensional, simply-rotating black hole on the brane \cite{DHKW,CKW,IOP}.
The resulting differential energy emission rate per unit
time, energy and angle of emission is given by the expression \cite{DHKW, CKW}
\begin{equation}
\frac{d^3E}{d(\cos\theta)\,dt\,d\omega}  =
\frac{1+\delta_{|s|,1}}{4\pi}\,
\sum _{l,m}
\frac{\omega}{\exp(\tilde \omega/T_H) \pm 1}
{\bf T}_\Lambda\,\left({}_{-h}S^2_\Lambda+{}_hS^2_\Lambda\right)\,.
\label{power-ang}
\end{equation}
The radiation spectrum of the black hole resembles those of a black body
with a temperature
\begin{equation}
T_H=\frac{(n+1) + (n-1) a_*^2}{4\pi\,(1+a_*^2)\,r_h}\,.
\label{T_H}
\end{equation}
At the same time, however, the spectrum is significantly modified compared to
the black-body one: in the exponent, the combination $\tilde \omega=\omega-am/(a^2+r_h^2)$,
includes the effect of the rotation of the black hole; also, the quantity
${\bf T}_\Lambda$, the transmission
probability (or, greybody factor), determines the number of particles that
eventually overcome the gravitational barrier of the black hole and reach asymptotic
infinity. If Eq. (\ref{power-ang}) is integrated over all angles of emission
$\theta$, we obtain the power rate in terms of unit time and energy
\begin{equation}
\frac{d^2E}{dt\,d\omega}=\frac{1+\delta_{|s|,1}}{2\pi}\,
\sum_{l=|s|}^{\infty}\,\sum_{m=-l}^{+l}\,\frac{\omega}{\exp(\tilde \omega/T_H) \pm 1}\,
{\bf T}_\Lambda\,.
\label{power-int}
\end{equation}

The derivation of the integrated-over-all-angles power spectra, for all species
of brane-localised fields -- scalars, fermions and gauge bosons, was performed
both analytically \cite{CEKT2,CEKT3} and numerically \cite{DHKW, CKW, IOP}. 
According to these results, the energy emission rate -- as well as the particle
and angular-momentum emission rates -- are significantly enhanced as both the
number of additional, spacelike dimensions and the angular-momentum of the
black hole increase. The enhancement factor was of order ${\cal O}(100)$ when
$n$ varied between 1 and 6, and of order ${\cal O}(10)$ as $a_*$ increased from
zero towards its maximum value $a_*^{max}=(n+2)/2$. 

In contrast to the case of the spherically-symmetric Schwarzschild phase, the
emission of particles during the rotating phase of the life of the black hole 
is not isotropic. The axis of rotation introduces a preferred direction in
space and the emitted radiation exhibits an angular variation as $\theta$
ranges from 0 to $\pi$. It was found \cite{DHKW, CKW, IOP} that a centrifugal
force is exerted on all species of particles, that becomes stronger as either
$\omega$ or $a$ increases and forces the particles to be emitted along the
equatorial plane ($\theta=\pi/2$). In addition, for particles with 
non-vanishing spin, an additional force, sourced by the spin-rotation coupling,
aligns the emitted particles parallel or antiparallel to the rotation axis
of the black hole -- this effect is more dominant the smaller the energy
and larger the spin of the particle is. If the form (\ref{power-ang}) of
the power spectrum is used where both helicities appear, the spectrum is
symmetric over the two hemispheres, $\theta \in [0, \pi/2]$ and [$\pi/2, \pi$].
If a modified form, in which only one of the helicities appear each time,
is used instead, then the angular profile is asymmetric with particles with
positive helicity (corresponding to ${}_{-|s|}S^2_\Lambda$) being emitted
in the upper hemisphere and particles of negative helicity (corresponding
to ${}_{+|s|}S^2_\Lambda$) being emitted in the lower one. This angular
variation in the Hawking radiation spectra is considered to be one of the main
observable effects on the brane of a higher-dimensional, rotating, decaying
black hole.

One would ideally like to deduce the values of both spacetime parameters, $n$
and $a$, from the predicted forms of the Hawking radiation spectra. However,
the fact that both parameters affect the 
integrated-over-all-angles spectra in a similar way impose a great obstacle.
The resolution of this problem would demand the existence of an observable
that depends strongly on only one of the two parameters while being (almost)
insensitive to the other. That observable was shown \cite{FST, CDKW3} to
be the angular variation of the spectra discussed above. Particularly, in the
low-energy channel, the alignment of the gauge bosons along the rotation
axis can reveal the orientation of the angular-momentum of the black hole.
Then, it was demonstrated that the angle of emission of fermions, in the same
energy channel, is very sensitive to the value of the angular-momentum
of the black hole: the larger the $a$ parameter is, the larger the value of
$\theta$, around which the emission is peaked, becomes. Remarkably, the behaviour
of gauge bosons and fermions alike remains unaltered as the dimensionality of
spacetime changes. 


\section{Analytical forms of the radial and angular functions
\label{sec-solutions}}

The results on the angular profile of the emitted fields with non-zero spin
on the brane, discussed above, were derived by numerically integrating both the
radial (\ref{radial}) and the angular (\ref{angular}) equation: the latter
in order to find the angular eigenvalue ${}_hA_\Lambda$ and eigenfunction
${}_{h}S_\Lambda$, and the former in order to determine the greybody factor
${\bf T}_\Lambda$ through the radial function ${}_hR_\Lambda$. The numerical
manipulation of the radial and angular differential equations is necessary
for the derivation of the exact solutions for ${}_hR_\Lambda$ and ${}_{h}S_\Lambda$,
respectively, and subsequently of the complete Hawking radiation spectra.
However, when it comes to the spectra of gauge bosons and fermions revealing
information about the orientation of axis and value of the angular momentum
of the black hole, the range of interest is the low-energy one. Thus, in what
follows we will focus on the low-energy channel, and attempt to derive analytically
information about the angular profile of non-zero-spin fields emitted on the brane.
To this end, we will henceforth ignore the single-component scalar fields and
concentrate our study on brane-localised fields with spin $1/2$ and $1$.

Under the assumption of low-energy of the emitted field and low-angular-momentum
of the black hole, the radial equation (\ref{radial}) was analytically solved
in \cite{CEKT2, CEKT3} for all species of particles. A well-known approximation
method was used
in which the radial equation was solved first near the horizon, then at asymptotic
infinity, and the two were finally matched at an intermediate regime to construct
the complete solution for ${}_hR_\Lambda$. The transmission probability 
${\bf T}_\Lambda$ for fermions was defined as the ratio of the flux of particles
at the black-hole horizon over the one at infinity, with the flux being determined
through the conserved particle current. For gauge bosons, where no conserved
particle current exists, a radial function redefinition and a simultaneous
change of the radial coordinate conveniently change the corresponding gravitational
potential to a short-range one - then, the amplitudes of the outgoing and incoming
plane waves at infinity can easily determine the transmission probability. 
For fermions and gauge bosons, ${\bf T}_\Lambda$ comes out to have the form
\cite{CEKT3}
\beq
{\bf T}^{(1/2)}_\Lambda = 1 - {{4\omega ^2 } \over {{}_\frac{1}{2}A_\Lambda
+1+a^2\omega^2 }}\left| {{{Y_{\frac{1}{2}} ^{(out)} } \over
{Y_{\frac{1}{2}} ^{(in)} }}} \right|^2 \label{grey-fer}\eeq 
and
\beq
{\bf T}^{(1)}_\Lambda  = 1 - \frac{16\omega ^4}{({}_1A_\Lambda+2+a^2\omega^2)^2}
\left|\frac{Y_1^{(out)}}{Y_1^{(in)}}\right|^2\,, \label{grey-gb}\eeq
respectively, where
\beq
\frac{Y_h^{(out)}}{Y_h^{(in)}}=\frac{\Gamma \left( {1 + Z } \right)\,
\Gamma \left( {{1 \over 2} + h + {Z \over 2} } \right)}
{(2i\omega)^{2h}\,\Gamma \left( {{1 \over 2} - h + {Z\over 2} } \right)\,
\left[\Gamma \left( {1 + Z } \right)\,e^{i\pi \left({1 \over 2} - h + {Z\over 2}\right)}
+B\,\Gamma \left( {{1 \over 2} + h + {Z \over 2} } \right)\right]}\,.
\eeq
and 
\begin{equation} B \equiv {{\Gamma (Z )} \over {\Gamma ({1 \over 2} - h
+\frac{Z}{2} ) }}\,{{\Gamma (c - a - b)\Gamma (a)\Gamma (b)}
\over {\Gamma (c - a)\Gamma (c - b)\Gamma (a + b - c)}}\,
\frac{\left[ {(1 + a_* ^2 )\,r_h^{n +
1} } \right]^{2\beta + |s| + B_*  - 2}}{(2i\omega )^{Z}}
\label{B1B2}\eeq
In the above, the quantity $Z$, defined by 
\beq Z=\sqrt{(2|s|-1)^2+4({}_hA_{\Lambda}+2|s|+a^2\omega^2)}\,,
\label{zeta}\eeq
appears in the solution of the radial equation in the asymptotic infinity that
is expressed in terms of the Kummer functions $M$ and $U$. Similarly, the
coefficients ($a$, $b$, $c$), given by
\begin{eqnarray}
a=\alpha + \beta +D_*-1\,, \qquad b=\alpha + \beta\,, \qquad c=1
-|s| + 2 \alpha\,,
\end{eqnarray}
are the coefficients of the hypergeometric function $F$ in terms of which the
solution of the radial equation is written near the black-hole horizon.
Finally, the following definitions hold \cite{CEKT3}
\begin{equation}
D_* \equiv 1 -|s| + \frac{2|s|+n\,(1+a_*^2)}{A_*} - \frac{4
a_*^2}{A_*^2}\,, \qquad 
\alpha   = {{|s|} \over 2} - \left( {{{iK_* } \over {A_* }} + {h
\over 2}} \right)\,, \label{sol-a}
\end{equation}
\begin{equation}
\beta =\frac{1}{2}\,\biggl[\,(2-|s|-D_*) - \sqrt{(D_*+|s|
-2)^2 - \frac{4 K_*^2-4ihK_*A_*}{A_*^2} -  \frac{4(4ih\omega_*-
{}_h\tilde\lambda_{\Lambda})\,(1+a_*^2)} {A_*^2}}\,\biggr]\,. \label{beta}
\end{equation}
supplemented by the following ones: $A_*  = n + 1 + (n - 1)a_*^2$,
$K_*  = K/r_h$ and
${}_h\tilde\lambda_{\Lambda}={}_h\lambda_{\Lambda} + 2|s|$. 

For scalar fields, the transmission probability is again defined from the
amplitudes of the outgoing and ingoing spherical waves at infinity \cite{CEKT2}
\bea
{\bf T}^{(0)}_\Lambda = 1-\left|\frac{B-i}{B+i}\right|^2 = 
\frac{2i\left(B^*-B\right)}{B B^* + i\left(B^*-B\right)+1}\,, \label{grey-sc}
\eea
where $B$ now is given by the expression
\bea
&~& \hspace*{-1.8cm}B = -\frac{1}{\pi}
\frac{Z\,2^{2l}}{\left[\omega r_h\,(1+a_*^2)^\frac{1}{n+1}\right]^{2l+1}}
\,\frac{\Gamma^2\left(Z/2\right)
\Gamma(\alpha+\beta + D_* -1)\,\Gamma(\alpha+\beta)\,
\Gamma(2-2\beta - D_*)}{\Gamma(2\beta + D_*-2)\,
\Gamma(2+\alpha -\beta - D_*)\,\Gamma(1+\alpha-\beta)} \,. \label{Beq}
\eea

We note that the angular eigenvalue ${}_hA_{\Lambda}$ makes its appearance 
in the above analytic results both in Eq. (\ref{zeta}) and
Eq. (\ref{beta}). As already mentioned in the previous section, in the 
case of a rotating black hole, this quantity does not exist in closed form.
For arbitrary large values of the energy of the emitted particle and
angular momentum of the black hole, its value can be determined only via
numerical means - that was the method applied in \cite{DHKW, CKW} where
the complete spectra for scalars, fermions and gauge bosons were derived.
However, for low $\omega$ and low $a$, the angular eigenvalue of the
spin-weighted spheroidal harmonics can be expressed as a power series with
respect to $a \omega$ \cite{press1, fackerell, churilov, Seidel, BCC}
\beq
{}_hA_{\Lambda}= \sum _{k=0}^\infty \,f_k\,(a \omega)^k\,.
\label{power-series}
\eeq
By using the above power-series form for the angular eigenvalue and keeping
terms up to fourth order, the analytically derived formulae for the 
transmission probabilities (\ref{grey-fer}) and (\ref{grey-gb}) for fermions
and gauge bosons - as well as the one for scalar fields  - were shown in
\cite{CEKT2, CEKT3} to be in excellent agreement with the exact numerical ones
derived in \cite{DHKW, CKW}. The power-series expansion of the angular
eigenvalue is quite cumbersome and, up to the sixth order, can be found in
\cite{press1, fackerell, churilov, Seidel, BCC}. It is worth giving here,
some particularly simple formulae we have derived, for the needs of our
analysis, for the eigenvalues of fermions and gauge bosons up to second order,
namely
\beq
{}_{\frac{1}{2}}A_{\Lambda}=l(l+1)-\frac{3}{4}-\frac{m\,(a \omega)}{2l(l+1)}+
\left\{\frac{{\cal A}^2_{1/2} +{\cal B}^2_{1/2}}{2l (l+1)}-\frac{1}{2}+
\frac{m^2}{8}\,\left[\frac{1}{(l+1)^3} -\frac{1}{l^3}\right]\right\}(a \omega)^2+...\,
\label{eigen-fer}
\eeq
and
\bea
{}_{1}A_{\Lambda}&=&l(l+1)-2 -\frac{2m\,(a \omega)}{l(l+1)}+
\left\{2({\cal A}^2_1 +{\cal B}^2_1)\left[1-\frac{3}{l (l+1)}\right]\right.\nonumber\\[1mm]
&-& \left. 2m^2\left[\frac{3(l+2)}{(l+1)^3} -\frac{2l+3}{l^3 (l+1)^2}\right]
+(3-2l-2l^2)\right\}\frac{(a \omega)^2}{(2l+3)(2l-1)}+...\,,
\label{eigen-gb}
\eea
where
\beq
{\cal A}_h ={\rm max}\,(|m|,|s|)\,, \qquad 
{\cal B}_h =\frac{mh}{{\rm max}\,(|m|,|s|)}\,.
\eeq
In the above, we have given the values of the angular eigenvalues for
the positive helicities $h=1/2$ and $h=1$, respectively. The angular eigenvalues
exhibit a well-known symmetry \cite{Leaver, BCC} according to which, if
${}_{|s|}A_\Lambda$ is the eigenvalue for the positive-helicity component of
a given field, then the one for the negative helicity ${}_{-|s|}A_\Lambda$ readily
follows from the relation ${}_{-|s|}A_\Lambda={}_{|s|}A_\Lambda+2|s|$.
For completeness, we add here a similar formula for the angular eigenvalue
of scalar fields that first appeared in \cite{Harris}:
\beq{}_{0}A_{\Lambda}=l(l+1)+ \left[\frac{ \displaystyle 1-2l-2(l^2-m^2)}
{\displaystyle (2l-1)\,(2l+3)}\right](a \omega)^2+...\,.
\label{eigen-sc}
\eeq

Let us now turn to the angular equation (\ref{angular}). Leaver \cite{Leaver} found
an analytic solution for the angular eigenfunction ${}_hS_\Lambda(x)$ that
may be expressed as a series of the following form
\beq
{}_hS_\Lambda(x) =e^{a\omega x}\,(1+x)^{k_-}\,(1-x)^{k_+}\,
\sum_{p=0}^\infty \,a_p\,(1+x)^p\,,
\label{ang-Leaver}
\eeq
where $x =\cos\theta$ and $k_{\pm} \equiv |m\pm h|/2$. The expansion
coefficients $a_p$ can be found through a three-term recursion relation
\beq
\alpha_0 a_1 +\beta_0 a_0=0\,, \label{a_1}
\eeq
\beq
\alpha_p\,a_{p+1} + \beta_p\,a_p + \gamma_p\,a_{p-1}=0\,, \qquad (p=1,2,...)
\label{a_p}
\eeq
In the above, the coefficients ($\alpha_p$, $\beta_p$, $\gamma_p$) are in 
turn determined by the relations
\bea
\alpha_p &=& -2(p+1)(p+2 k_-+1)\,,\nonumber \\[2mm]
\beta_p &=& p(p-1) +2 p(k_-+k_++1-2a \omega) -[a^2\omega^2+h(h+1) +{}_hA_\Lambda]
\nonumber \\[1mm]
&-&[2 a\omega (2k_-+h+1)-(k_-+k_+)(k_-+k_++1)]\,, \label{coefficients}\\[2mm]
\gamma_p &=& 2a \omega (p+k_-+k_++h)\,.\nonumber
\eea
The above analytic form determines the angular eigenfunction up to a
constant that can be fixed by imposing the normalization condition
$\int_{-1}^1 |{}_hS_\Lambda(x)|^2 dx=1$. According to \cite{BCC},
an excellent approximation to the exact solution is obtained by keeping
$\sim 10$ terms in the expansion of (\ref{ang-Leaver}).


\section{Analytical description of the angular profile
\label{sec-profile}}

In this section, we will attempt to study the angular profile of the emitted
Hawking radiation on the brane by employing semi-analytic techniques. Our
starting point will be Eq. (\ref{power-ang}) that determines the angular
profile of the emitted radiation as a function of $x=\cos\theta$. By using
the analytical formulae presented in the previous section, we will compute
the value of the angle $\theta$ where the emission of particles becomes
maximum. Since the emission of positive and negative helicity components
is symmetric under the change $\theta \rightarrow \pi-\theta$ \cite{CDKW3},
in what follows we consider only the emission of positive helicity components, $h>0$. 

In Eq. (\ref{power-ang}), the dependence on the angle $\theta$ is restricted
in the angular eigenfunction ${}_hS_\Lambda(x)$. One may then naively try to
extremize this equation to find a constraint that will determine the desired
value of $\theta_{\rm max}$, defined as the value of the angle where the
differential rate of emission takes its maximum value. We then obtain
\begin{equation}
\frac{d\,\,}{dx}\left(\frac{d^3E}{dx\,dt\,d\omega}\right)  =
\frac{1+\delta_{|s|,1}}{4\pi}\,
\sum _{l,m}
\frac{\omega}{\exp(\tilde \omega/T_H) \pm 1}
{\bf T}_\Lambda \left(2\,{}_hS_\Lambda \frac{d {}_hS_\Lambda}{dx}\right)=0\,.
\label{extremum-0}
\end{equation}
By employing the analytical expression (\ref{ang-Leaver}) for the angular
eigenfunction ${}_hS_\Lambda(x)$ and evaluating the derivative, we
obtain the following constraint
\begin{equation}
\sum _{l,m} {}_hW_\Lambda
(1+x)^{2k_-}\,(1-x)^{2k_+}\,
\sum_{p=0}^\infty a_p\,(1+x)^p\,\sum_{q=0}^\infty a_q\,(1+x)^q
\left(a \omega +\frac{k_-+q}{1+x}-\frac{k_+}{1-x}\right)=0\,.
\label{extremum-1}
\end{equation}
In the above, we have defined the ``weight factor'' ${}_hW_\Lambda$ as
\beq
{}_hW_\Lambda \equiv \frac{\omega}{\exp[(\omega-m\Omega)/T_H) \pm 1}\,{\bf T}_\Lambda\,.
\label{weight}
\eeq

The analytical evaluation of the constraint (\ref{extremum-1}) in full is not possible.
As mentioned above, the sum over $p$ (and $q$), originating from the analytic
form of the angular eigenfunction, may be truncated at a finite value, but
care must be taken so that the truncated series remains close to the exact
solution and the value of $\theta_{\rm max}$ is not affected. The constraint contains
two additional sums: one with respect to $l$, the total angular-momentum
number ranging from $|s|$ to $\infty$, and one over $m$, the azimuthal
angular-momentum number that takes values in the range $[-l,+l]$. None of
these sums can be discarded: all of the quantities involved, the coefficients
$k_\pm$, $a_p$ (and $a_q$), as well as the weight factor ${}_hW_\Lambda$,
depend on both angular-momentum numbers in a non-trivial manner. It is,
therefore, the combined contributions of all, in principle, partial modes 
that determines the angular profile of the emitted radiation. Finally, these
contributions do not enter on an equal footing: each mode carries a weight
factor ${}_hW_\Lambda$ -- defined in Eq. (\ref{weight}) in terms of the
`thermal/statistics' function and the greybody factor ${\bf T}_\Lambda$ --
that determines the magnitude of its contribution to the angular profile. 

In what follows, we will attempt to shed light to the important contributions 
to Eq. (\ref{extremum-1}) that determine the value and location, in terms
of the angle $\theta$, of the maximum emission rate for fermions and gauge
bosons. As the interesting phenomena take place in the low-energy regime,
we will use purely analytic expressions for all quantities involved, namely
the angular eigenvalue,  the angular eigenfunctions and the greybody factor.
Having been established in the literature \cite{CDKW3} that the orientation
of the emission of fermions and gauge bosons is not affected by the value
of the number of extra dimensions introduced in the model, we will keep
fixed the value of $n$ and, henceforth, set $n=2$.

\subsection{Emission of Fermions}
\label{fermions}

We will start with the most phenomenologically interesting case, the emission
of fermions. Our strategy will be the following: by using the most complete
analytic forms, we will investigate when a particular contribution to the
angular profile becomes so small that is irrelevant and can thus be ignored.
We will therefore use the power series expansion (\ref{power-series}) for the
angular eigenvalue up to fourth order in $(a \omega)$, the analytic form of
the angular eigenfunction given in (\ref{ang-Leaver}) by keeping
terms~\footnote{We have confirmed that, by
keeping terms up to $p=10$ in this expansion, the derived values of the angular
eigenfunction agree extremely well with the exact numerical ones -- as a
consistency check, we have successfully reproduced the plots of the angular
eigenfunctions appearing in \cite{BCC, CKW}.} up to $p=10$, and, at the beginning,
allow the angular-momentum numbers  ($l$, $m$) to vary over their full range. 

\begin{figure}[t]
\begin{center}
\mbox{
\includegraphics[width=8.4cm]{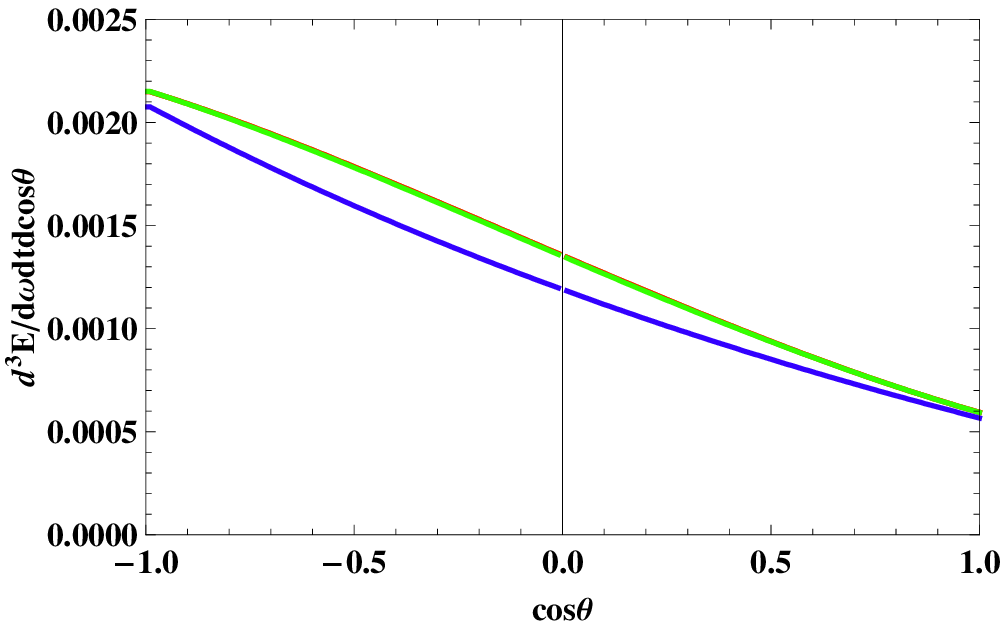}}\hspace*{-0.2cm}
\includegraphics[width=8.4cm]{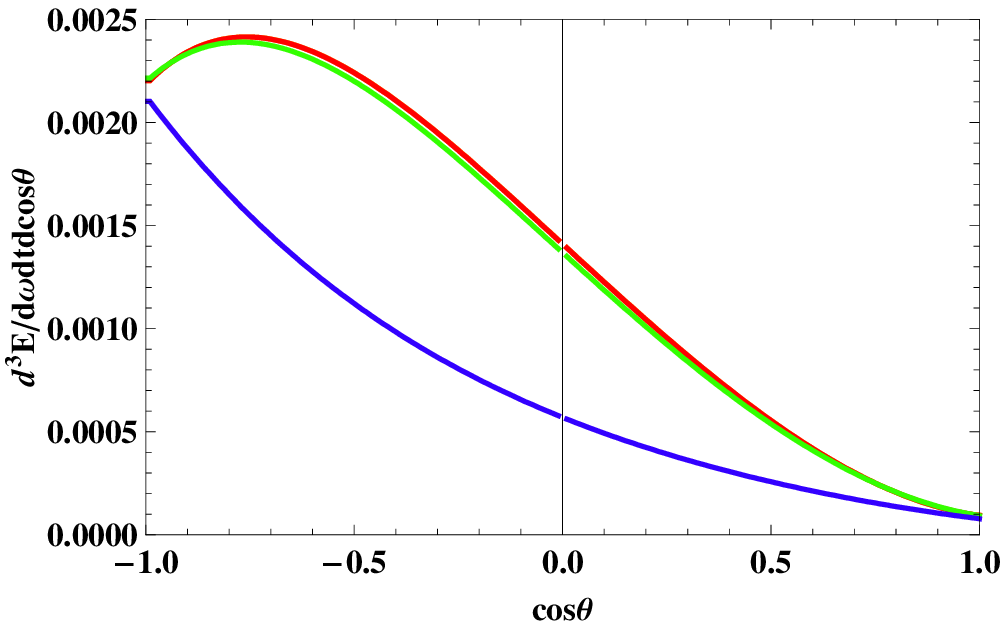}
\end{center}
\caption{The differential energy emission rate (\ref{power-ang}) in terms of
$\cos\theta$, for $n=2$, $\omega_*=0.5$, and (a) $a_*=0.5$ (left plot) and (b)
$a_*=1.5$ (right plot). The different curves correspond to the emission rate
when partial modes up to $l=1/2$, $l=3/2$, $l=5/2$ and $l=7/2$ (from bottom to top)
have been summed up.}
\label{max-l}
\end{figure}

In Figs. \ref{max-l}(ab), we depict the differential emission rate (\ref{power-ang})
per unit time, unit frequency and angle of emission in terms of $\cos\theta$,
for the case $\omega_*=0.5$ and $a_*=0.5$ (left plot) and $a_*=1.5$ (right plot).
The different curves correspond to the derived spectrum where modes up to a certain
value of $l$ (and all values of $m$ in the range $[-l,+l]$\,) have been summed up:
the lower (blue) curve includes only the $l=1/2$ modes, the next (green) one modes
up to $l=3/2$, the subsequent (red) one modes up to $l=5/2$ and the last (orange) one
modes up to $l=7/2$. We observe that the $l=7/2$ curve is not even visible as it
is completely covered by the $l=5/2$ one -- the same happens for all higher modes.
As a matter of fact, the difference between the $l=5/2$ and $l=3/2$ curves is also
quite small: for the maximum value of the angular momentum considered, $a_*=1.5$,
the difference in the value of the emission rate at its maximum and of
$\theta_{\rm max}$ is of the order of only 1\%; for smaller values of
$a_*$, the errors reduce even more: for $a_*=0.5$, the difference
in the value of the emission rate at its maximum drops at the level of 0.08\%
while $\theta_{\rm max}$ is not affected at
all. We may thus conclude, that the sum over $l$ in (\ref{extremum-1}) can be
safely truncated at $l=3/2$. The reason for this significant truncation is the
weight factor ${}_hW_\Lambda$: although the thermal/statistics factor gives a
boost to modes with large and positive $m$, the significant suppression of the 
greybody factor ${\bf T}_\Lambda$ in the low-energy regime as $l$ increases ensures
that higher modes can be safely ignored.

\begin{figure}[t]
\begin{center}
\includegraphics[width=10cm]{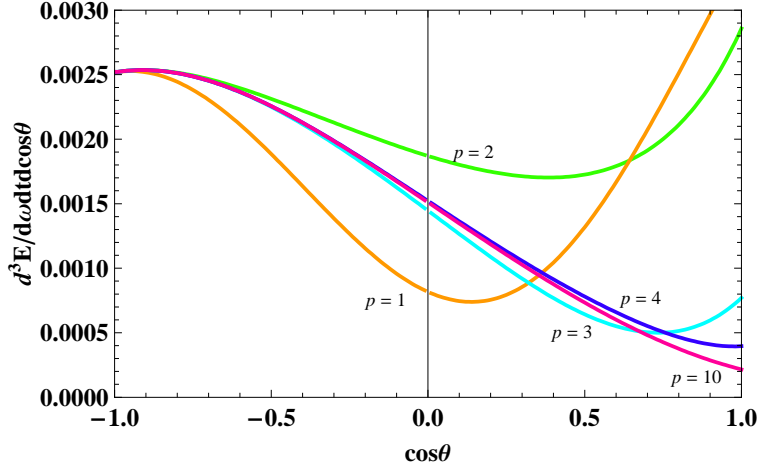}
\end{center}
\caption{Energy emission rate per unit time, unit frequency and angle of
emission in terms of $\cos\theta$, for $n=2$, $\omega_*=0.5$ and $a_*=1$,
and terms in the series expansion of the angular eigenfunction summed up
to $p=1$, $p=2$, $p=3$, $p=4$ and $p=10$.}
\label{max-Sp}
\end{figure}

As a next step in our study, we investigate whether the sum in the series
expansion of the eigenfunction can also be truncated. To this end, we have
computed the differential energy emission rate (\ref{power-ang}), for
$\omega_*=0.5$ and $a_*=1$, by keeping modes up to $l=5/2$ for extra safety,
and gradually increasing the maximum value of the sum index $p$. The
behaviour of the corresponding results for the emission rate as a function again
of  $\cos\theta$ is plotted in Fig. \ref{max-Sp}, where the different curves
correspond to the maximum value of $p$ kept in the sum, $p=1,2,3,4$ and 10.
We observe that the correct value of the emission rate at its maximum
is obtained fairly soon, when terms only up to $p=2$ are included in
the sum; the value of $\theta_{\rm max}$, on the other hand, needs one
more term in the expansion ($p=3$) to acquire its actual value. Our results are
not in contradiction with \cite{BCC} where the value of $p=10$ was defined
as the one that accurately reproduces the exact form of the eigenfunction.
Indeed, higher terms included in the sum up to $p=10$ do change the behaviour
of the eigenfunction, however, these changes are restricted in the area 
away from the angle of maximum emission, as Fig. \ref{max-Sp} clearly shows.
The value of the angular momentum of the black hole strongly affects the 
value of $p_{\rm max}$: for $a_*=1.5$, the correct value of $\theta$ is obtained when
terms up to $p=4$ are included; in contrast,  for $a_*=0.5$, no terms higher
than $p=1$ are needed in the sum.

\begin{figure}[t]
\begin{center}
\mbox{
\includegraphics[width=8cm]{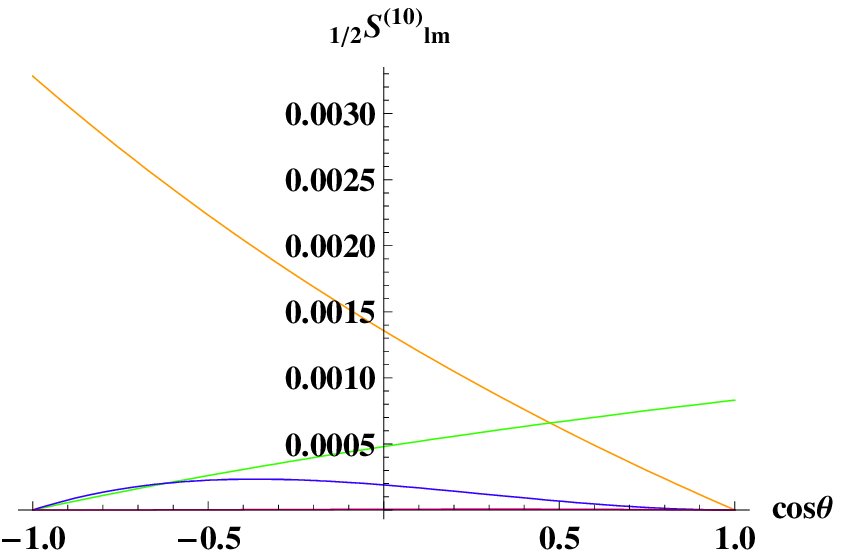}}
\includegraphics[width=8cm]{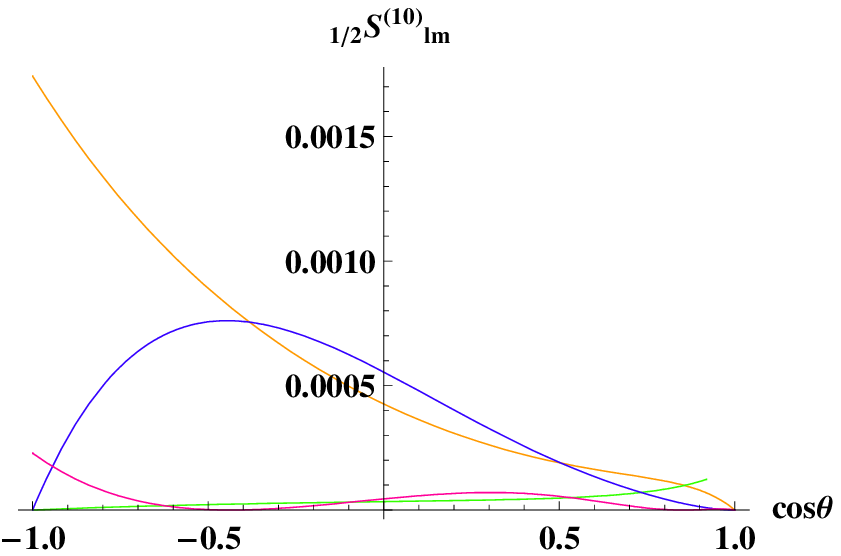}
\end{center}
\caption{The fermionic angular eigenfunction $_{1/2}S^{(10)}_{lm}$ 
as a function of $\cos\theta$, for $n=2$, $\omega_*=0.5$ and: a) $a_*=0.5$
(left plot) and $(l,m)=[(\frac{1}{2},\frac{1}{2}), (\frac{1}{2},-\frac{1}{2}),
(\frac{3}{2},\frac{3}{2}), (\frac{3}{2},\frac{1}{2})]$ (from top to bottom),
b) $a_*=1.5$ (right plot) and $(l,m)=[(\frac{1}{2},\frac{1}{2}), (\frac{3}{2},\frac{3}{2}),
(\frac{3}{2},\frac{1}{2}), (\frac{1}{2},-\frac{1}{2})]$ (from top to bottom).}
\label{dominant_modes}
\end{figure}

Let us comment at this point on the expression of the angular eigenvalue that
was used in our calculations. As noted above, we initially employed the power
series form of Eq. (\ref{power-series}) with terms up to the fourth order
in $(a \omega)$. However, we have found that the expression (\ref{eigen-fer}),
with terms up to second order only, is more than adequate to lead to 
accurate results. Although including higher-order terms cause, at times,
a significant change in the value of the angular eigenvalue itself, that
change hardly affects any aspects of the angular profile of the emitted radiation.
For example, for the mode $l=1/2$ and $m=-1/2$, the difference in the
value of the eigenvalue, when terms up to second and third order, respectively,
are kept, is of the order of 10\%, the effect in the value of the coefficient
$\beta_p$ appearing in Eq. (\ref{coefficients}) is only 0.2\% which
leaves the angular profile virtually unchanged. 

One may simplify further the analysis by considering more carefully the
partial modes that dominate the energy emission spectrum. According to the
results above, the sum over $l$ can be safely truncated at the value
$l=3/2$, and thus we need to sum over the following six modes:
$(l,m)=[(\frac{1}{2},\frac{1}{2}), (\frac{1}{2},-\frac{1}{2}),
(\frac{3}{2},\frac{3}{2}), (\frac{3}{2},\frac{1}{2}), (\frac{3}{2},-\frac{1}{2}),
(\frac{3}{2},-\frac{3}{2})]$. However, not all of the above modes have
the same contribution to the angular variation of the energy spectrum.
In Fig. \ref{dominant_modes}(a), we display the angular eigenfunctions of the
four most dominant modes out of the aforementioned six, for $n=2$,
$\omega_*=0.5$ and angular momentum $a_*=0.5$ (left plot). It is clear that,
for small values of $a_*$, the two $l=1/2$ modes dominate over the
$l=3/2$ ones. This dominance is further enhanced when the corresponding
weight factors are taken into account, with the ones for the $l=3/2$ modes
being at least one order of magnitude smaller than the ones for the $l=1/2$ modes. 
But even the contribution of the two dominant modes, $(\frac{1}{2},\pm\frac{1}{2})$,
is not of the same magnitude: when the weight factors and the difference
in magnitude of the angular eigenfunctions are taken into account, the
$(\frac{1}{2},\frac{1}{2})$ mode is found to have at least five times
bigger contribution than the $(\frac{1}{2},-\frac{1}{2})$ one. As a result,
the angular pattern of the emitted radiation at the low-energy channel,
for small values of the angular momentum parameter, is predominantly defined
by the $(\frac{1}{2},\frac{1}{2})$ mode. Then, the constraint (\ref{extremum-1})
takes the simplified form~\footnote{In what follows, we will adopt the value
$p=3$ as the maximum value of the sum index needed to accurately reproduce
the behavior of the fermionic eigenfunction around the angle of maximum emission.}
\beq
\sum_{q=0}^3 a_q\,(1+x)^q
\left(a \omega +\frac{q}{1+x}-\frac{1}{2(1-x)}\right)=0\,,
\eeq
and more particularly
\bea 
&& \hspace*{-1cm}
\left(a\omega-\frac{1}{2(1-x)}\right) \left[a_0^{(1/2)} +a_1^{(1/2)} (1+x) +
a_2^{(1/2)} (1+x)^2+
a_3^{(1/2)} (1+x)^3\right]\nonumber \\[2mm] && \hspace*{6cm}
+a_1^{(1/2)}+2a_2^{(1/2)} (1+x) +3a_3^{(1/2)} (1+x)^2=0\,.
\label{extr_m05}
\eea
In the above, we have used that $k_-=0$ and $k_+=1/2$ for the mode
$(\frac{1}{2},\frac{1}{2})$, and the superscript $\{1/2\}$ denotes that the 
set of coefficients $a_p^{(1/2)}$ for this particular mode should be used here.
In Appendix \ref{app_fermions}, we list the
results for the angular eigenvalue, as this follows from Eq. (\ref{eigen-fer}),
the values of the  $(\alpha_p, \beta_p, \gamma_p)$ coefficients, according
to the definitions (\ref{coefficients}), and finally the relations between
the first four sum coefficients $a_p$, given by the three-term recursion
relations (\ref{a_1})-(\ref{a_p}). 
A simple numerical analysis, then, shows that Eq. (\ref{extr_m05}) does
not have any roots in the range $x \in (-1,+1)$ for $a \omega < 0.52$, 
with the global maximum located at $x=-1$ and the global minimum at $x=1$.
Therefore, if we fix the energy channel at e.g. $\omega_*=0.5$, the
angular eigenfunction of the $(\frac{1}{2},\frac{1}{2})$-mode does not show
any extrema up to $a_* \simeq 1$; as a result, the energy emission rate
takes its maximum value at $\theta=\pi$ in accordance with the exact
numerical results derived in \cite{FST, CDKW3}.

Nevertheless, as $a_*$ increases, the $(\frac{3}{2},\frac{3}{2})$-mode
becomes important -- this may be clearly seen in Fig. \ref{dominant_modes}(b).
Let us examine the behaviour of this mode on its own. Its extremization
constraint is given now by 
\bea
&& \hspace*{-1cm}
\left(a\omega+\frac{1}{2(1+x)}-\frac{1}{(1-x)}\right) \left[a_0^{(3/2)} +
a_1^{(3/2)} (1+x) +a_2^{(3/2)} (1+x)^2+ a_3^{(3/2)} (1+x)^3\right]\nonumber
\\[2mm] && \hspace*{6cm}+a_1^{(3/2)}+2a_2^{(3/2)} (1+x) +3a_3^{(3/2)} (1+x)^2=0\,,
\label{extr_m3p2}
\eea
where we have used that, for this mode, $k_-=1/2$ and $k_+=1$. By making use of the 
relations between the first four $a_p^{(3/2)}$ coefficients, as these are found again in
Appendix \ref{app_fermions}, and performing a simple numerical analysis, we
arrive at the following results: 
for $a\omega=0$, all $a_i$ with $i\geq 1$ vanish, and the constraint 
(\ref{extr_m3p2}) reveals the existence of a sole extremal point at
$x=-1/3$; this extremum is a local maximum -- as $a_*$ increases, the
local maximum becomes gradually more important and slowly moves to the left,
thus competing with the maximum of the $(\frac{1}{2},\frac{1}{2})$-mode
at $x=-1$ to create a global maximum for the energy emission rate in the
range $(-1,-1/3)$ with the exact location depending on the value of $a_*$. 

\begin{figure}[t]
\begin{center}
\includegraphics[width=8cm]{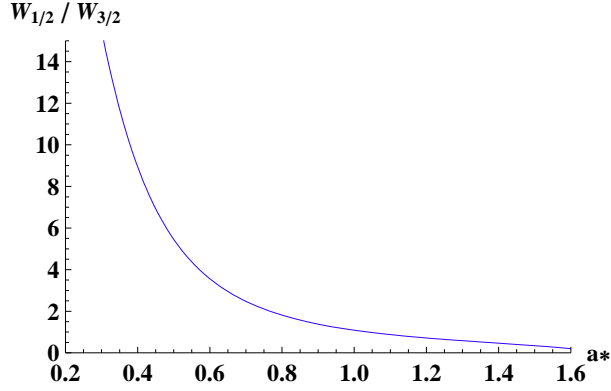}
\end{center}
\caption{The relative weight factor $W_{rel}=W_{1/2}/W_{3/2}$ in terms of the
angular-momentum parameter $a_*$, for the particular case of $\omega_*=0.6$.}
\label{weight-factor}
\end{figure}

Thus, summarizing the above results, for an arbitrary value of $a_*$, the angular
variation of the emitted fermions is mainly determined by the contribution of
the $(\frac{1}{2},\frac{1}{2})$ and $(\frac{3}{2},\frac{3}{2})$ modes, and
thus the constraint (\ref{extremum-1}) may take the final form
\bea
&&{}W_{rel}\,
\sum_{p=0}^3 a_p^{(1/2)}\,(1+x)^p\,\sum_{q=0}^3 a_q^{(1/2)}\,(1+x)^q
\left(a \omega +\frac{q}{1+x}-\frac{1}{2(1-x)}\right)\nonumber \\
&& \hspace*{0.5cm}+
(1-x^2) \sum_{p=0}^3 a_p^{(3/2)}\,(1+x)^p\,\sum_{q=0}^3 a_q^{(3/2)}\,(1+x)^q
\left(a \omega +\frac{1/2+q}{1+x}-\frac{1}{1-x}\right)=0\,.
\label{extremum-2}
\eea
We have also defined the relative ``weight factor''
$W_{rel}=W_{1/2}/W_{3/2}$ whose value depends strongly on the angular 
parameter $a_*$ -- this dependence is shown in Fig. \ref{weight-factor}.
For small values of $a_*$, $W_{rel}$ takes large values and the extremization
constraint is dominated by the $(\frac{1}{2},\frac{1}{2})$-mode causing the
emitted fermions to be aligned with the rotation axis. As $a_*$ increases,
$W_{rel}$ decreases reaching the value one for approximately $a_*=1$ --
now, both modes contribute equally and $\theta_{\rm max}$ is pushed away 
from the $\theta=\pi$ value. For even larger values of $a_*$, the 
$(\frac{3}{2},\frac{3}{2})$-mode starts dominating with the angle
of maximum emission moving further away.

In support of our argument, that the $(\frac{1}{2},\frac{1}{2})$ and
$(\frac{3}{2},\frac{3}{2})$ modes predominantly determine the angular variation
of the fermionic spectrum, in Table \ref{lpmax-fer} we display the
values of the energy emission rate at the angle of maximum emission
as well as the value of the corresponding angle $\theta_{\rm max}$, for
various values of the energy parameter $\omega_*$ and angular-momentum
parameter $a_*$. In each case, we display two values: the first one 
follows by taking into account the contribution of the two aforementioned
modes and keeping terms only up to $p=3$ in the sum of the angular
eigenfunction (or, up to $p=4$ for $a_* \geq 1$); the second follows by
keeping all terms up to $p=10$
and all partial modes up to $l=7/2$. The values of the energy parameter
$\omega_*$ have been chosen to lie in the low-energy regime and, at
the same time, to display a non-trivial angular variation of the spectrum
- it is worth mentioning that for all values smaller than $\omega_*= 0.5$, 
the angle of maximum emission is constantly located at $\theta=\pi$. On
the other hand, the angular-momentum parameter $a_*$ scans a fairly broad
range from $a_*=0.5$ to $a_*=1.5$.

For the energy channel $\omega_*=0.5$, the agreement between the two sets
of results is extremely good: the error in the value of the energy emission
rate at its maximum reaches the magnitude of 3.5\% at most, while the agreement
in the value of $\theta_{\rm max}$ is perfect. In agreement with the
exact numerical results \cite{CDKW3} where this energy channel was studied,
for small values of $a_*$, the emitted radiation remains very close to the
rotation axis and only for values close to $a_*=1.0$ the emission starts
showing a maximum at a gradually smaller angle. For $\omega_*=0.6$,
the errors in the value of the emission rate and $\theta_{\rm max}$ are  
at the level of 5\% and 3\% respectively, with the emission being peaked
at an angle away from the horizon axis for $a_* \geq 0.75$. For $\omega_*=0.7$,
the error in the value of $\theta_{\rm max}$ is still quite small~\footnote{The error
in the value of $\theta_{\rm max}$ is indeed quite small for all values of
$a_* \geq 0.75$. For $a_*=0.5$, we observe a significant deviation of $\theta_{\rm max}$
from its actual value for the energy channels $\omega_*=0.7$ and $\omega_*=0.8$.
This is due to the fact that, for these specific values of the energy parameter
and angular momentum , the mode $(\frac{1}{2}, -\frac{1}{2})$ that we have ignored
in our approximation is of the same order of magnitude as the $(\frac{3}{2}, \frac{3}{2})$ that
we have taken into account.}
ranging between 3\% and 4\%, whereas the error in the value of the emission rate
at its maximum is now taking large values (7\%-17\%).  
Finally, for completeness, we show the energy channel of $\omega_*=0.8$:
although we have probably exceeded the range of validity of our
approximation, the error in the value of $\theta_{\rm max}$ remains less than 10\%. 

The above comparison demonstrates that, for low values of the parameters
$\omega_*$ and $a_*$ where our semi-analytic approximation is valid, the use
of the two modes, the $(\frac{1}{2},\frac{1}{2})$ and $(\frac{3}{2},\frac{3}{2})$
ones, and the constraint (\ref{extremum-2}) can provide realistic results
for the angular variation of the fermionic spectrum. This can consequently
help to determine the value of the angular momentum of the black hole
according to the proposal of \cite{FST, CDKW3}. The
results displayed in Table \ref{lpmax-fer} confirm the behaviour found
numerically for the energy channel $\omega_*=0.5$ \cite{CDKW3}, extend the
set of values that could be used for comparison with experiment to additional
low-energy values of $\omega_*$ and, finally, provide a very satisfactory
semi-analytic approximation in terms of only two partial modes. 

\begin{table}
\begin{center}
$\begin{array}{|c||c|c|c|c|} \hline & \omega_*=0.5 & \omega_*=0.6 & \omega_*=0.7
& \omega_*=0.8 \\ \hline \hline
&\hspace*{-0.3cm}\begin{array}{cc}{\rm approx.} & {\rm \quad full}\end{array}&
\hspace*{-0.3cm}\begin{array}{cc}
{\rm approx.}& {\rm \quad full}\end{array}&\hspace*{-0.3cm}\begin{array}{cc}
{\rm approx.}& {\rm \quad full}\end{array}&\hspace*{-0.3cm}\begin{array}{cc}{\rm approx.}&
{\rm \quad full}\end{array}
\\ \hline a_*=0.50 & \begin{array}{cc} \begin{array}{c}  2.078 \\ -0.99 \end{array}
 & \begin{array}{c} 2.150 \\ -0.99 \end{array} \end{array} &
\begin{array}{cc} \begin{array}{c} 1.956 \\ -0.99 \end{array} &
\begin{array}{c} 2.053 \\ -0.99 \end{array} \end{array} &
\begin{array}{cc} \begin{array}{c}  1.545 \\ -0.99 \end{array}
 & \begin{array}{c}  1.657 \\ -0.89 \end{array} \end{array} &
\begin{array}{cc} \begin{array}{c}  1.028 \\ -0.87 \end{array}
 & \begin{array}{c}  1.132 \\ -0.74 \end{array} \end{array}
\\ \hline a_*=0.75 & \begin{array}{cc} \begin{array}{c}  2.316 \\ -0.99 \end{array}
 & \begin{array}{c}  2.406 \\ -0.99 \end{array} \end{array} &
\begin{array}{cc} \begin{array}{c} 2.054 \\ -0.82 \end{array} &
\begin{array}{c} 2.168 \\ -0.81 \end{array} \end{array} &
\begin{array}{cc} \begin{array}{c}  1.693 \\ -0.66 \end{array}
 & \begin{array}{c}  1.827 \\ -0.64 \end{array} \end{array} &
\begin{array}{cc} \begin{array}{c}  1.232 \\ -0.57 \end{array}
 & \begin{array}{c}  1.402 \\ -0.52 \end{array} \end{array}
\\ \hline a_*=1.00 & \begin{array}{cc} \begin{array}{c}  2.444 \\ -0.91 \end{array}
 & \begin{array}{c}  2.535 \\ -0.91 \end{array} \end{array} &
\begin{array}{cc} \begin{array}{c} 2.170 \\ -0.68 \end{array} &
\begin{array}{c} 2.282 \\ -0.67 \end{array} \end{array} &
\begin{array}{cc} \begin{array}{c}  1.809 \\ -0.56 \end{array}
 & \begin{array}{c} 1.970  \\ -0.54 \end{array} \end{array} &
\begin{array}{cc} \begin{array}{c}  1.303 \\ -0.51 \end{array}
 & \begin{array}{c}  1.552 \\ -0.46 \end{array} \end{array}
\\ \hline a_*=1.25 & \begin{array}{cc} \begin{array}{c}  2.488 \\ -0.81 \end{array}
 & \begin{array}{c}  2.567 \\ -0.81 \end{array} \end{array} &
\begin{array}{cc} \begin{array}{c} 2.090 \\ -0.61 \end{array} &
\begin{array}{c} 2.205 \\ -0.60 \end{array} \end{array} &
\begin{array}{cc} \begin{array}{c}  1.590 \\ -0.52 \end{array}
 & \begin{array}{c} 1.784  \\ -0.50\end{array} \end{array} &
\begin{array}{cc} \begin{array}{c}  1.019 \\ -0.46 \end{array}
 & \begin{array}{c}  1.338 \\ -0.43 \end{array} \end{array}
\\ \hline a_*=1.50 & \begin{array}{cc} \begin{array}{c}  2.347 \\ -0.76 \end{array}
 & \begin{array}{c}  2.415 \\ -0.76 \end{array} \end{array} &
\begin{array}{cc} \begin{array}{c} 1.715 \\ -0.57 \end{array} &
\begin{array}{c} 1.831 \\ -0.56 \end{array} \end{array} &
\begin{array}{cc} \begin{array}{c}  1.038 \\ -0.41 \end{array}
 & \begin{array}{c}  1.248\\ -0.42 \end{array} \end{array} &
\begin{array}{cc} \hspace*{0.2cm}\begin{array}{c} - \\ - \end{array}
 & \begin{array}{c}  \quad\,\, 0.098 \\ \quad\,\, -0.40 \end{array} \end{array}
\\ \hline \end{array}$
\end{center}
\caption{The approximated and full values of the energy emission rate
(\ref{power-ang}) at the angle of maximum emission, in units of $10^{-3}/r_h$,
and the corresponding values of $\cos(\theta_{\rm max})$ for fermions. \label{lpmax-fer}}
\end{table}

\subsection{Emission of Gauge Bosons}
\label{gbosons}

Let us now address the emission of gauge bosons on the brane by the
simply-rotating black hole. We will again focus on the low-energy regime
as this is the energy channel at which the emission of gauge bosons is
polarised along the rotation axis of the black hole. We will attempt to
determine the main factors that contribute to this behaviour and, if
possible, provide analytical arguments that justify it.

Following a similar strategy as in the case of fermions, we first investigate
whether the infinite sum over the partial modes, characterised by ($l,m$),
in Eq. (\ref{extremum-1}) can be truncated. By gradually increasing the
value of $l$ (and summing over all corresponding values of $m$), we looked
for that value beyond which any increase in $l$ makes no difference to the
value of the energy emission rate at its maximum and of the corresponding
angle. It turns out that, at the low-energy regime, this value is reached
very quickly -- this behaviour is clearly displayed by the entries of Table
\ref{lpmax-gb}. In the upper part of the Table, we present the energy emission
rate (\ref{power-ang}) at its maximum and the corresponding angle as we
increase $l$ from 1 to 3 and vary $a_*$ from 0.5 to 1.5 in a random
low-energy channel ($\omega_*=0.3$). We observe that the value of the angle 
of maximum emission for positive-helicity ($h=1$) gauge bosons is indeed 
$\theta=\pi$, i.e. anti-parallel to the angular-momentum vector of the
black hole, and that this value is not affected at all by adding any partial
modes beyond the ones with $l=1$. The energy emission rate also varies very
little: its value at the angle of maximum emission is already reached for
$l=2$ and the difference from its value when only the $l=1$ modes are taken
into account is of the order of 0.1\% independently of the value of the
angular-momentum of the black hole. We may thus conclude that the angular
profile of the emission of gauge bosons at the low-energy regime is determined
almost exclusively by the lower $l=1$ modes: the sum over $l$, therefore, in
Eq. (\ref{extremum-1}) can be replaced by the contribution of only its first
term. 

\begin{table}
\begin{center}
$\begin{array}{|c|c|c|c|} \hline & a_*=0.5 & a_*=1.0 & a_*=1.5
\\ \hline \hline l_{\rm max} & \begin{array}{cc}  \quad {\rm Rate}
& \quad \,\,{\rm \cos(\theta_{max})}  \end{array} & \begin{array}{cc} \quad {\rm Rate}
& \quad \,\,{\rm \cos(\theta_{max})} \end{array} & \begin{array}{cc} \quad {\rm Rate}
& \quad \,\,{\rm \cos(\theta_{max})} \end{array}
\\ \hline l=1 & \begin{array}{cc} \hspace*{-0.5cm}0.017866 & \hspace*{0.3cm}
-0.99 \end{array} &
\begin{array}{cc} \hspace*{-0.5cm}0.028779 & \hspace*{0.3cm}-0.99 \end{array}
& \begin{array}{cc}
\hspace*{-0.5cm}0.066357 & \hspace*{0.3cm}-0.99 \end{array}
\\  l=2 & \begin{array}{cc} \hspace*{-0.5cm} 0.017883 & \hspace*{0.3cm}
-0.99  \end{array} &
\begin{array}{cc} \hspace*{-0.5cm}0.028817 & \hspace*{0.3cm}-0.99 \end{array}
& \begin{array}{cc}
\hspace*{-0.5cm}0.066440 & \hspace*{0.3cm}-0.99 \end{array}
\\  l=3 & \begin{array}{cc} \hspace*{-0.5cm}0.017883 & \hspace*{0.3cm}
-0.99 \end{array} &
\begin{array}{cc} \hspace*{-0.5cm}0.028817 & \hspace*{0.3cm}-0.99 \end{array}
& \begin{array}{cc}
\hspace*{-0.5cm}0.066440 & \hspace*{0.3cm}-0.99 \end{array}
\\ \hline \hline p_{\rm max} & \begin{array}{cc} \quad {\rm Rate}
& \quad \,\,{\rm \cos(\theta_{max})}  \end{array} & \begin{array}{cc} \quad {\rm Rate}
& \quad \,\,{\rm \cos(\theta_{max})} \end{array} & \begin{array}{cc} \quad {\rm Rate}
& \quad \,\,{\rm \cos(\theta_{max})} \end{array}
\\ \hline p=0 & \begin{array}{cc} \hspace*{-0.5cm}0.017948 &
\hspace*{0.3cm}-0.99 \end{array} &
\begin{array}{cc} \hspace*{-0.5cm}0.029051 & \hspace*{0.3cm}-0.99 \end{array}
& \begin{array}{cc}
\hspace*{-0.5cm}0.067319 & \hspace*{0.3cm}-0.99 \end{array}
\\  p=1 & \begin{array}{cc} \hspace*{-0.5cm}0.017866 &
\hspace*{0.3cm}-0.99  \end{array} &
\begin{array}{cc} \hspace*{-0.5cm}0.028778 & \hspace*{0.3cm}-0.99 \end{array}
& \begin{array}{cc}
\hspace*{-0.5cm}0.066353 & \hspace*{0.3cm}-0.99 \end{array}
\\  p=2 & \begin{array}{cc} \hspace*{-0.5cm}0.017866 &
\hspace*{0.3cm}-0.99 \end{array} &
\begin{array}{cc} \hspace*{-0.5cm}0.028779 & \hspace*{0.3cm}-0.99 \end{array}
& \begin{array}{cc}
\hspace*{-0.5cm}0.066357 & \hspace*{0.3cm}-0.99 \end{array}
\\  p=3 & \begin{array}{cc} \hspace*{-0.5cm}0.017866 &
\hspace*{0.3cm}-0.99  \end{array} &
\begin{array}{cc} \hspace*{-0.5cm}0.028779 & \hspace*{0.3cm}-0.99 \end{array}
& \begin{array}{cc}
\hspace*{-0.5cm}0.066357 & \hspace*{0.3cm}-0.99 \end{array}
\\ \hline \hline\end{array}$
\end{center}
\caption{The differential energy emission rates at the maximum angle of
emission and the corresponding angle for gauge bosons, for $\omega_*=0.3$ and $n=2$, in
terms of the angular-momentum number $l$ and sum index $p$, and for three
indicative values of $a_*$. \label{lpmax-gb}}
\end{table}

We performed a similar analysis regarding the value of $p$ in the sum in
the expression for the angular eigenfunction, and we have found similar
results displayed in the lower part of Table \ref{lpmax-gb}. The value of
the angle of maximum emission is again not affected as terms beyond the
first one ($p=0$) are added. The actual value of the energy emission rate
at the maximum angle is also very loosely dependent on $p$: as $p$ goes from
1 to 2, the difference is of the order of $10^{-3}\%$, while the difference
between the cases with $p=0$ and $p=1$ is again very small, of the order of
0.5\%. While, according to the above, the sum over $p$ can be clearly truncated
even at $p=0$, to increase the validity of the subsequent analysis, we will
also keep terms with $p=1$, and thus write the analytic expression 
(\ref{ang-Leaver}) of the angular eigenfunction as
\beq
{}_1S_\Lambda(x) =e^{a\omega x}\,(1+x)^{k_-}\,(1-x)^{k_+}\,
[a_0+a_1\,(1+x)]\,.
\label{ang-gb}
\eeq

A final point that needs to be addressed is the contribution of the different
$m$-modes. For $l=1$, we have three modes with $m=+1,0,-1$ that have, nevertheless,
a different weight factor and thus a different contribution to the constraint
(\ref{extremum-1}). A numerical evaluation of the weight factor (\ref{weight}),
with ${\bf T}_\Lambda$ given in Eq. (\ref{grey-gb}), for these three modes, in conjunction
with the value of the angular eigenfunction in each case, reveals that the contribution
of the $m=1$ mode to the constraint (\ref{extremum-1}) is almost two orders of
magnitude larger than the one of the $m=0$ mode, and that in turn is larger by 
two orders of magnitude than the contribution of the $m=-1$ mode. Therefore, it
is the $l=m=1$ mode that effectively determines the angular profile of the 
emitted radiation. 

Then, the constraint (\ref{extremum-1}) can take a particularly simple form.
For $l=m=1$ and $h=1$, we obtain $k_-=0$ and $k_+=1$, which then leads to the
condition 
\beq
a_0 \left(a \omega -\frac{1}{1-x}\right) + 
a_1\,(1+x)\left(a \omega +\frac{1}{1+x}-\frac{1}{1-x}\right)=0\,.
\label{extr-gb}
\eeq
The above can be written as a quadratic polynomial in $x$,
with solutions
\beq
x_{\rm ex}=-\left(\frac{a_0}{2a_1}+\frac{1}{a\omega}\right) \pm
\sqrt{1+\frac{1}{(a\omega)^2} + \frac{a_0}{a_1}+\frac{a_0^2}{4a_1^2}}\,.
\label{extrema-gb}
\eeq
If the above values correspond to extremal points in the regime $x \in (-1,1)$,
then they should satisfy the inequality $|x_{\rm ex}|<1$. This in turn imposes
constraints on the coefficients $a_0$ and $a_1$. As in the case of fermions, these
coefficients, for a given set of numbers ($h,l,m$), are given solely in terms
of the parameter $a \omega$. In Appendix \ref{app_gb}, we present the main
steps for the derivation of the relations between the sum coefficients $a_p$
in the case of gauge bosons.  There, it is found that, for the mode $h=l=m=1$,
\beq
\frac{a_1^{(1)}}{a_0^{(1)}}=-\frac{a \omega}{2}\,(3+\frac{9a\omega}{20})\,.
\eeq
We substitute the above ratio into Eq. (\ref{extrema-gb}), and demand
that $-1<x_{\rm ex}<1$. While the left-hand-side inequality is automatically satisfied
for all values of $a \omega$, the right-hand-side translates to 
$9 (a \omega)^2 +60 a\omega -20>0$ that leads to the constraint $a \omega > 0.32$. 

Therefore, for $a\omega < 0.32$ no extremal points for the differential energy
emission rate exist in the range $x \in (-1, 1)$. This quantity is thus monotonic
and has global extremal points at the end points $x=-1$ and $x=+1$. Substituting
$k_-=0$ and $k_+=1$ in Eq. (\ref{ang-gb}), it is easy to see that, for $x=+1$, 
the angular eigenfunction vanishes, while, for $x=-1$, it takes its maximum value
$2 a_0 e^{-a \omega}$. As a result, the positive-helicity component of the
gauge field is perfectly aligned in an anti-parallel direction to the angular-momentum
vector of the black hole ($\theta=\pi$), in agreement with the exact numerical
results \cite{CDKW3}. If $a \omega$ exceeds the value 0.32, a local maximum
develops at an internal point of the range $(-1, 1)$, however, this remains
subdominant to the global maximum at $x=-1$ up to the value $a \omega \simeq 0.85$.
Therefore, if we fix the energy channel to $\omega_* =0.5$, the maximum of the
emitted radiation in the form of gauge fields remains aligned in an antiparallel
direction to the angular-momentum of the black hole for all values of $a_*$ up
to 1.7, in agreement again with the exact numerical results \cite{CDKW3}. 


\subsection{Emission of Scalars}
\label{scalars}

We finally address the case of the emission of scalar fields on the brane
by a simply-rotating black hole. Although no useful information regarding
the angular momentum of the black hole can be derived in this case, for
completeness, we briefly discuss the main characteristics of the angular
pattern of the scalar emission and the main contributing factors. 

In order to investigate whether it is possible again to truncate the sums over
$l$ and $p$, that appear in the constraint (\ref{extremum-1}), we construct
Table \ref{lpmax-sc}. The left-hand-side of the table displays the energy
emission rate at the angle of maximum emission and the corresponding angle
in terms of the angular-momentum number $l$. The energy channel $\omega_*=0.3$
has been chosen as an indicative case, the number of extra dimensions has
been again fixed to $n=2$, and the angular-momentum parameter is taken to
be $a_*=1.5$ -- this is the highest value of $a_*$ considered in this analysis,
and the one for which the convergence of the sums over $l$ and $p$ is 
the most difficult to achieve. We observe that all modes beyond $l=2$ add
a contribution of order 0.01\%, and thus can be safely ignored. But the difference
between the values of the emission rate when all modes up to $l=1$ and $l=2$
have been, respectively, summed up is also very small, of the order of 1\%.
The value of $\theta_{\rm max}$ has also been stabilised to $\pi/2$ when $l_{max}=1$.
Therefore, in the context of our semi-analytic approach, the sum over $l$ can
be indeed truncated at $l=1$ with no significant error. 

On the right-hand-side of Table \ref{lpmax-sc}, we keep all partial modes up
to $l=2$ for extra accuracy, and examine the convergence of the sum over $p$.
The change in the value of the energy emission rate at the angle of maximum
emission between the cases with $p=4$ and $p=5$ is of the order of 0.3\%,
while all higher contributions are an order of magnitude smaller. The value
of $\theta_{\rm max}$ has also taken the exact value of $\pi/2$, therefore this
sum can be safely truncated at $p=4$. 

\begin{table}
\begin{center}
$\begin{array}{|c|l||c|l|} \hline \hline l_{\rm max} & \begin{array}{cc}  \quad {\rm Rate}
& \quad \,\,{\rm \cos(\theta_{max})}  \end{array} & p_{\rm max} & \begin{array}{cc}
\quad {\rm Rate} & \quad \,\,{\rm \cos(\theta_{max})} \end{array}
\\ \hline l=0 & \begin{array}{cc} 0.00052329 & \hspace*{0.3cm}\pm 1 \end{array} &
p=1 & \begin{array}{cc} 0.00438439 & \hspace*{0.3cm}\pm 1 \end{array}
\\  l=1 & \begin{array}{cc} 0.00137162 & \hspace*{0.6cm}0  \end{array} &
p=2 & \begin{array}{cc} 0.00165378 & \hspace*{0.3cm}\pm 0.67 \end{array}
\\  l=2 & \begin{array}{cc} 0.00139658 & \hspace*{0.6cm}0 \end{array} &
p=3 & \begin{array}{cc} 0.00135215 & \hspace*{0.3cm}\pm 0.05 \end{array}
\\  l=3 & \begin{array}{cc} 0.00139688 & \hspace*{0.6cm}0 \end{array} &
p=4 & \begin{array}{cc} 0.00137489 & \hspace*{0.6cm}0 \end{array}
\\  l=4 & \begin{array}{cc} 0.00139688 & \hspace*{0.6cm}0 \end{array} &
p=5 & \begin{array}{cc} 0.00137122 & \hspace*{0.6cm}0\end{array}
\\  l=5 & \begin{array}{cc} 0.00139688 & \hspace*{0.6cm}0 \end{array} &
p=6 & \begin{array}{cc} 0.00137167 & \hspace*{0.6cm}0 \end{array}
\\ \hline \hline\end{array}$
\end{center}
\caption{The differential energy emission rates at the maximum angle of
emission and the corresponding angle, for $\omega_*=0.3$, $n=2$ and $a_*=1.5$,
in terms of the angular-momentum number $l$ and sum index $p$ for scalar
fields. \label{lpmax-sc}}
\end{table}

An exhaustive analysis of the values of the weight factors of the contributing
partial modes $(l,m)=\{(0,0),(1,-1),(1,0),(1,1)\}$ for a variety of energy
channels, $\omega_* \in(0.2-0.8)$ and angular-momentum of the black hole,
$a_* \in (0.5-1.5)$, reveals that the two most dominant modes are the
$(0,0)$ and $(1,1)$ with the contributions of the other two being always two
orders of magnitude smaller. Therefore, combining all the above results,
the extremization constraint (\ref{extremum-1}) for the case of scalar
fields, takes the simplified form:
\bea
&&\hspace*{-0.5cm}W_{rel}\,
\sum_{p=0}^4 a_p^{(00)}\,(1+x)^p\,\sum_{q=0}^4 a_q^{(00)}\,(1+x)^q
\left(a \omega +\frac{q}{1+x}\right)\nonumber \\
&& \hspace*{0.5cm}+
\sqrt{1-x^2}\,\sum_{p=0}^4 a_p^{(11)}\,(1+x)^p\,\sum_{q=0}^4 a_q^{(11)}\,(1+x)^q
\left(a \omega +\frac{1/2+q}{1+x}-\frac{1}{2(1-x)}\right)=0\,.
\label{extremum-sc}
\eea
In the above, we have used that for the $(0,0)$-mode, $k_-=k_+=0$, while
for the $(1,1)$-mode, $k_-=k_+=1/2$. Also, in this case, the relative weight
factor is defined as $W_{rel} \equiv W_{00}/W_{11}$. The expressions of the
sum coefficients $a_p^{(00)}$ and $a_p^{(11)}$ for the two modes can be
found at the Appendix  \ref{app_sc}.

Let us consider individually the two dominant modes. Starting from the
$l=m=0$ mode, we write its extremization constraint as
\beq
\sum_{p=0}^3 \,[ a\omega\,a_p + (p+1)\,a_{p+1}\,]\,(1+x)^p +
a \omega \,a_4 (1+x)^4=0\,. \label{polyn-00}
\eeq
This is a polynomial of fourth degree that in principle has four
roots and, therefore, four potential extremal points. However, if we use
the expressions of the $a_p$ coefficients for the $l=m=0$ mode listed
in Appendix \ref{app_sc}, we find that two of these roots are complex
conjugates and one lies outside the range $[-1,1]$. Thus, the angular
wavefunction of the $l=m=0$ mode has only one extremal point with respect
to $x=\cos\theta$. This extremum is a minimum located at $x=0$ ($\theta=\pi/2$)
for small values of $a \omega$ that moves to positive values of $x$ as
$a \omega$ increases. However, the latter effect is actually an artifact
of the truncation of the sum in the expression of the angular eigenfunction
at a finite value of $p$. Even in our approximation where terms up to $p=4$
are kept, we may see that the constant term of the polynomial (\ref{polyn-00})
is given by a particular combination of the $a_p$ coefficients that due to 
multiple cancellations quickly tends to zero, namely
\beq
\sum_{p=0}^3 \,[ a\omega\,a_p + (p+1)\,a_{p+1}\,] +
a \omega \,a_4 \simeq -\frac{(a \omega)^4}{72}+ {\cal O} (a \omega)^5\,. 
\label{const-00}
\eeq
Had we kept all terms in the series expansions of the angular eigenvalue
and eigenfunction, every subsequent term in the sum of (\ref{const-00})
would cancel part of the remain of all previous ones all the way to infinity,
thus ensuring that the $x=0$ is always an extremum of the $l=m=0$ mode. A
simple numerical analysis then shows that this local extremum is the only one
in the range $(-1,1)$ and corresponds to a minimum. Due to the fact that
$k_+=k_-=0$, the $l=m=0$ mode reaches the same maximum value at the boundary
points $x=\pm 1$. 

Moving to the next dominant mode $l=m=1$, its extremization constraint reads
\beq
\sum_{p=0}^4 \,a_p\,(1+x)^{p} \left[a\omega\,(1-x^2)-x (p+1)+p\right]=0\,.
\label{polyn-11}
\eeq
This is a polynomial of sixth degree whose six roots are potential extremal
points. Substituting the $a_p$ coefficients for this mode from Appendix
\ref{app_sc} and performing a simple numerical analysis, one may see that
the four roots are two pairs of complex conjugate numbers and one lies outside
the range $[-1,1]$ leaving again only one root that may indeed correspond to
a local extremal point of the angular eigenfunction of the $l=m=1$ scalar
mode with respect to $x=\cos\theta$. As in the case of the $l=m=0$ mode,
the extremum is located at $x=0$ and moves towards positive values of $x$
as the parameter $a\omega$ increases. We have again confirmed that the constant
term of the above polynomial tends again to zero very quickly, i.e.
\beq
\sum_{p=0}^4 \left(a\omega +p\right) a_p
\simeq -\frac{(a \omega)^4}{375}+ {\cal O} (a \omega)^5\,,
\label{const-011}
\eeq
signalling the fact that the $x=0$ is always an extremum of the angular
eigenfunction of the $l=m=1$ mode. The difference from the case of the
$l=m=0$ mode lies in the fact that now this extremum is a global maximum
instead of a minimum with the angular eigenfunction of the $l=m=1$ mode
vanishing at the boundary points $x=\pm 1$ since $k_+=k_-=1/2$. Let us
briefly add here that a similar analysis of the remaining two scalar modes,
$l=1,m=0$ and $l=-m=1$, shows that these follow the behaviour of the
$l=m=0$ and $l=m=1$ modes, respectively.

The exact numerical analysis of the emission of scalar fields on the brane
by a simply-rotating higher-dimensional black hole \cite{DHKW} has revealed
that the corresponding spectrum shows no angular variation for low values
of the energy parameter $\omega_*$ and of the angular-momentum number $a_*$.
Clearly, for $a_* =0$, the constraint (\ref{polyn-00}) is trivially satisfied 
and the $l=m=0$ mode shows no extremal points -- note that, for the mode
$l=m=1$, the constraint (\ref{polyn-11}) still leads to a maximum at $x=0$
even at $a_*=0$. For low values of $\omega_*$, a careful analysis reveals
that it is the $l=m=0$ mode that dominates over the others, therefore, for
low $a_*$, the spectrum remains spherically-symmetric. As $a_*$ starts
increasing,  the $l=m=0$ also develops an extremum at $x=0$ -- it turns out
that there is always a low-energy regime where the minima of the $l=m=0$
and $l=1,m=0$ modes exactly cancel the maxima of the $l=m=1$ and $l=-m=1$ 
modes leading again to a spherically symmetric spectrum, however, this
energy regime becomes gradually more narrow. If we allow the energy parameter
$\omega_*$ to increase, too, then fairly quickly the $l=m=1$ mode starts
dominating causing the spectrum to exhibit maximum emission at $x=0$,
i.e. on the equatorial plane ($\theta=\pi/2$), in agreement with the
exact numerical results \cite{DHKW}.
 

\section{Discussion and conclusions
\label{sec-conclusions}}

One of the most exciting prospects of the theories predicting the existence
of additional spacelike dimensions in nature and a low fundamental scale
for gravity is the potential creation of higher-dimensional black holes
from the collision of ordinary brane-localised particles. 
If the scale is low enough, the creation could in principle take place
at ground-based particle accelerators and possibly be observed in the
near future. The main observable signal is considered to be the emission
of Hawking radiation on the brane in the form of ordinary Standard Model
particles. 

During the study of the spherically-symmetric Schwarzschild phase, it was
found that the radiation spectra of higher-dimensional black holes -- even
if we focus on the part of the emission that takes place on the brane
where ourselves, the observers, are located -- show a strong dependence on
the number of additional spacelike dimensions that exist transversely to
our brane. Therefore, the expectation was formed that the detection of
the Hawking radiation spectra could lead to the determination of the
number of extra spacelike dimensions in nature. 
However, if the angular-momentum of the black hole is taken
into account -- which generically is non-zero and seems to dominate almost
all of the life of the black hole -- this dependence on $n$ is entangled
with the dependence on the angular-momentum parameter $a$. 

During the spin-down phase of the life of the produced black
hole, the emission exhibits, among other features, a strong angular
variation in the radiation spectra with respect to the rotation axis
of the black hole. It has been suggested \cite{FST, CDKW3} that this angular
variation is the observable that could disentangle the dependence of the
radiation spectra on $n$
and $a$ as it depends strongly on the latter while being (almost)
insensitive to the former. It was found that, in the low-energy channel,
the emitted gauge bosons become aligned to the rotation axis of the
produced black hole while fermions form an angle with the rotation axis
whose exact value depends on the angular-momentum of the black hole. 

Attacking the problem of the angular variation of the Hawking radiation
spectra in an exact way, and for all values of the parameters of the theory,
is extremely challenging. It demands the numerical determination of both the
radial and angular eigenfunction of the emitted fields as well as the numerical
calculation of the angular eigenvalue that connects the corresponding
equations. In addition, the angular pattern of the emitted spectra is formed
from the contribution of an, in principle, infinite number of partial modes,
numbered by the pair of angular-momentum numbers $(l,m)$, each entering in
the expression of the emission rate with its own weight (thermal and
greybody) factor. Therefore, the use of the formal extremization constraint,
that we have derived and which should determine the angles of maximum
emission of all species of particles, seems rather unrealistic.

Nevertheless, as the exact numerical analyses \cite{FST, CDKW3} have shown,
all the valuable information that we should deduce from the angular spectra
are restricted in the low-energy regime. In this regime, one may use
approximate techniques to solve the radial equation and thus determine the
weight-factor of each contributing partial mode \cite{CEKT2, CEKT3}. In addition,
analytic formulae for the angular eigenvalue and eigenfunction exist 
\cite{press1, fackerell, churilov, Seidel, BCC, Leaver} that allow us to
study the problem of the angular 
variation of the spectra without resorting to complex numerical techniques.
Combining the above tools in a constructive but critical manner, we were
able to study the angular variation of the Hawking radiation spectra of
fermions, gauge bosons and scalar fields in a semi-analytic way. 

Starting  from the case of fermionic fields, the use of the analytic form of the
greybody factor allowed us to compute the weight factor of each contributing
partial mode. This, combined with a power series form for both the angular
eigenvalue and eigenfunction, led to the isolation of the partial modes
that predominantly determine the angular pattern of the corresponding
radiation spectra in the low-energy regime. Also, by demanding that the
errors associated to the elimination of all higher-order terms were small,
we were able to truncate the infinite sums in both the expressions of the
angular eigenvalue and eigenfunction. At the end, we demonstrated that
the contribution of only two partial modes, the $(\frac{1}{2},\frac{1}{2})$
and $(\frac{3}{2},\frac{3}{2})$, was more than adequate to provide approximate
results for the value of the angle of maximum emission and of the corresponding
emission rate that were within a range of 5\% accuracy of the full results.
Our study was completed by the derivation of the values of the above quantities,
both in an exact and approximate way, for a variety of values of the energy
parameter $\omega_*$ of the emitted fermionic field and angular-momentum
parameter $a_*$ of the black hole, that could in principle be used for the
determination of the angular momentum upon the observation of such a radiation
spectrum. 

Whereas the angular variation of the radiation spectra of the emitted fermions
is very sensitive to the value of the angular-momentum of the black hole -- the
larger the $a$ parameter is, the larger the value of $\theta_{\rm max}$ -- the
orientation of the gauge bosons emitted in the low-energy regime was found, by
the exact numerical analyses, to be constantly aligned to the rotation axis of
the black hole. Thus although it seems that no further information can be deduced
from the study of the gauge bosons, we nevertheless performed the same analysis
in an attempt to justify analytically the predicted behaviour. We demonstrated,
by using a similar strategy as in the case of fermions, that a single mode,
the $l=m=1$, mainly determines the angular profile of the emitted gauge bosons.
As its angular eigenfunction exhibits no extremal points up to $a \omega = 0.32$,
and even then these extrema remain subdominant to the global maximum at $\theta=0,\pi$
(for helicities $h=\mp 1$, respectively) up to $a \omega =0.85$, it is thus
confirmed that the emission of gauge bosons
in the low-energy regime will remain aligned to the rotation axis of the black
hole for a wide range of the angular-momentum parameter. For example, gauge bosons
emitted in the energy channel $\omega_*=0.5$ will remain mostly parallel or 
antiparallel to the rotation axis up to the value of $a_*=1.7$, however, they
will start deviating significantly from this behaviour for values of the angular
momentum parameter of the black hole larger than this.

For completeness, we have finally studied the case of the emission of scalar
fields on the brane by a simply-rotating black hole. In this case, the exact
numerical analyses have shown that the emission remains spherically-symmetric
for low values of $\omega$ and $a$ and then, as either of the two parameters
increases, the emission starts concentrating on the equatorial plane. One
could thus assume that by looking at the emission of scalar fields, the 
equatorial plane, and thus the rotation axis of the black hole, could again
be determined. Our analysis has shown that the angular profile of the
radiation spectra of scalars in the low-energy regime is mainly determined
by two modes, the $l=m=0$ and $l=m=1$, the first having a minimum at $\theta=\pi/2$
and the second a maximum at the same point. For small values of $\omega_*$
and $a_*$, we have confirmed that the combination of these two modes
creates indeed a ``spherically-symmetric zone" in the emission  where the
two extrema exactly
cancel each other. As either $\omega_*$ and $a_*$ increases further, it
is the $l=m=1$ mode that starts dominating pushing the bulk of the emitted
scalars towards the equatorial plane. Nevertheless, this transition becomes
gradually and is finally realised for values of the parameters of the theory
where our approximate techniques are not valid any more.

Closing, let us note that our analysis in this work was based on the study of
aspects, such as the existence of extremal points and relative magnitudes, of the
spin-weighted spheroidal harmonics. These functions arise in a variety of
problems, both in four dimensions as well as in the context of brane models, whenever
the study of spin-$s$ fields in a 4-dimensional spacetime with one angular-momentum
component is performed. We thus envisage that the properties of the angular
eigenfunctions revealed in this analysis as well as the analytic expressions of
the angular eigenvalues and $a_p$ coefficients for fermions, gauge bosons and
scalar fields presented in the Appendix will be of use in a variety of problems.
Nevertheless, let us stress that the aspects of the particular problem 
studied here, i.e. the angular profile of the emission of Standard Model
particles on the brane by a higher-dimensional black hole, could not be
performed only by means of 4-dimensional tools: the particles emitted
propagate on a brane embedded in a higher-dimensional spacetime, and this
is reflected in the expressions of the greybody factors that determine
to a great extent the weight factors of the individual partial modes.
It is therefore the combination of both traditional 4-dimensional and brane
techniques that has allowed us to analytically reproduce the angular
distribution of energy emission, and hopefully provide the means for the
determination of the angular momentum and axis of rotation of the
produced black hole.

\section*{Acknowledgements}
The authors would like to thank Marc Casals for his valuable help with the
normalization factors of the spin-weighted spheroidal harmonics.
This research has been co-financed by the European Union (European Social Fund - ESF)
and Greek national funds through the Operational Program "Education and Lifelong
Learning" of the National Strategic Reference Framework (NSRF) - Research Funding
Program: THALIS. Investing in the society of knowledge through the European Social Fund.

\appendix

\section{Expansion Coefficients for the Spin-Weighted Spheroidal Harmonics}

The three-term recursion relations (\ref{a_1})-(\ref{a_p}) can be used to determine
the coefficients $a_p$ that appear in the expansion of the spin-weighted
spheroidal harmonics $_{h}S_\Lambda$. For instance, for the first five expansion
coefficients, we obtain
\beq
a_1=-\frac{\beta_0}{\alpha_0}\,a_0\,, \qquad 
a_2=-\frac{\beta_1}{\alpha_1}\,a_1 - \frac{\gamma_1}{\alpha_1}\,a_0=
\left(\frac{\beta_1 \beta_0}{\alpha_1 \alpha_0}- \frac{\gamma_1}{\alpha_1}\right)a_0\,,
\eeq
\beq
a_3 =-\frac{\beta_2}{\alpha_2}\,a_2 - \frac{\gamma_2}{\alpha_2}\,a_1=
\left(-\frac{\beta_2 \beta_1 \beta_0}{\alpha_2 \alpha_1 \alpha_0}+
\frac{\beta_2 \gamma_1}{\alpha_2 \alpha_1} +\frac{\gamma_2 \beta_0}{\alpha_2 \alpha_0}
\right)a_0\,,
\eeq
\smallskip
\beq
a_4 =-\frac{\beta_3}{\alpha_3}\,a_3 - \frac{\gamma_3}{\alpha_3}\,a_2=
\left(\frac{\beta_3\beta_2 \beta_1 \beta_0}{\alpha_3\alpha_2 \alpha_1 \alpha_0}-
\frac{\beta_3 \beta_2 \gamma_1}{\alpha_3 \alpha_2 \alpha_1} -
\frac{\beta_3 \gamma_2 \beta_0}{\alpha_3 \alpha_2 \alpha_0} -
\frac{\gamma_3 \beta_1 \beta_0}{\alpha_3 \alpha_1 \alpha_0}+ 
\frac{\gamma_3 \gamma_1}{\alpha_3 \alpha_1}\right)a_0\,.
\eeq
The $(\alpha_p, \beta_p, \gamma_p)$ coefficients are given by Eqs. (\ref{coefficients})
and must be evaluated for each specific partial mode. 

\subsection{Fermions}
\label{app_fermions}

For the needs of our analysis, we determine the above coefficients for the
fermionic modes $(\frac{1}{2},\frac{1}{2})$ and $(\frac{3}{2},\frac{3}{2})$.
First, for the $(\frac{1}{2},\frac{1}{2})$-mode, Eq. (\ref{eigen-fer}) leads
to the following result for the corresponding eigenvalue ($h=1/2$)
\beq
_{\frac{1}{2}}A_{\frac{1}{2}\frac{1}{2}}=-\frac{a \omega}{3}-\frac{11}{27}\,(a \omega)^2\,,
\eeq
where we have used that, for this mode, $k_-=0$ and $k_+=1/2$. Then, the
$(\alpha_p, \beta_p, \gamma_p)$ coefficients take the form
\beq
\alpha_p^{(1/2)}=-2(p+1)^2\,, \quad \beta_p^{(1/2)}=p\,(p+2)-4a\omega\,(p+\frac{2}{3})-
\frac{16}{27}\,(a\omega)^2\,, \quad \gamma_p^{(1/2)}=2a\omega(p+1)\,.
\eeq
Then, the recursion relations (\ref{a_1})-(\ref{a_p}) lead to the following
relations between the first four sum coefficients
\beq
\frac{a_1^{(1/2)}}{a_0^{(1/2)}}=-\frac{4}{3}\,a\omega \left(1+\frac{2a\omega}{9}\right)\,,
\qquad
\frac{a_2^{(1/2)}}{a_0^{(1/2)}}=(a\omega)^2\left(1 + \frac{28 a\omega}{81}\right) +...\,,
\eeq
\smallskip
\beq
\frac{a_3^{(1/2)}}{a_0^{(1/2)}}=-\frac{392}{729}\,(a\omega)^3 \left(1 +
\frac{187}{441}\,a\omega\right)+...\,.
\label{a_1/2}
\eeq
The superscript $(1/2)$ denotes that the above expressions hold for the 
case of the $(\frac{1}{2},\frac{1}{2})$ partial mode. We have also given
only the relations between the first four expansion coefficients since, as shown
in section \ref{fermions}, in the case of fermions, it suffices to consider only
terms up to $p=3$ in the sum of Eq. (\ref{ang-Leaver}). 

For the mode $(\frac{3}{2},\frac{3}{2})$, using the fact that now $k_-=1/2$
and $k_+=1$, we arrive at the following result for the angular eigenvalue
\beq
_{\frac{1}{2}}A_{\frac{3}{2}\frac{3}{2}}=3-\frac{a \omega}{5}-
\frac{29}{125}\,(a \omega)^2\,.
\eeq
In turn, the $(\alpha_p, \beta_p, \gamma_p)$ coefficients take the form
\beq
\alpha_p^{(3/2)}=-2(p+1)(p+2)\,, \quad \beta_p^{(3/2)}=p(p+4)-4a\omega (p+\frac{6}{5})-
\frac{96}{125}\,(a\omega)^2\,, \quad \gamma_p^{(3/2)}=2a\omega(p+2)\,.
\eeq
Finally, the above result into the following relations between the first four
sum coefficients
\beq
\frac{a_1^{(3/2)}}{a_0^{(3/2)}}=-\frac{6}{5}\,a\omega \left(1+\frac{4a\omega}{25}\right)\,,
\qquad
\frac{a_2^{(3/2)}}{a_0^{(3/2)}}=\frac{4}{5}\,(a\omega)^2\left(1 + 
\frac{34 a\omega}{125}\right) +...\,,
\eeq
\smallskip
\beq
\frac{a_3^{(3/2)}}{a_0^{(3/2)}}=-\frac{716}{1875}\,(a\omega)^3 \left(1 +
\frac{1588}{4475}\,a\omega\right)+...\,.
\label{a_3/2}
\eeq

\subsection{Gauge Bosons}
\label{app_gb}

We now turn to the case of gauge bosons, and more particularly to
the dominant mode with $h=l=m=1$. Employing Eq. (\ref{eigen-gb}) and 
(\ref{coefficients}), we find the following results for the angular eigenvalue
\beq
_{1}A_{11}=-a \omega- \frac{11}{20}\,(a \omega)^2\,,
\eeq
and the $(\alpha_p, \beta_p, \gamma_p)$ coefficients
\beq
\alpha_p=-2 (p+1)^2\,, \qquad \beta_p=p(p+3) -a\omega(4p+3)-\frac{9(a\omega)^2}{20}\,,
\qquad \gamma_p=2a\omega (p+2)\,.
\eeq
We may then compute the relations between the different sum coefficients
$a_p$ - although for our analysis we need only the relation between $a_1$ and $a_0$,
for completeness, we display again the relations between the first four
coefficients
\beq
\frac{a_1^{(1)}}{a_0^{(1)}}=-\frac{a \omega}{2}\,(3+\frac{9a\omega}{20})\,,
\qquad
\frac{a_2^{(1)}}{a_0^{(1)}}=(a\omega)^2\left(\frac{6}{5} + 
\frac{9 a\omega}{32}\right) +...\,,
\eeq
\smallskip
\beq
\frac{a_3^{(1)}}{a_0^{(1)}}=-\frac{(a\omega)^3}{32}\left(\frac{65}{3} +
\frac{1247}{200}\,a\omega\right)+...\,.
\label{a_11}
\eeq

\subsection{Scalar Fields}
\label{app_sc}

We finally study the case of scalar fields ($h=0$). For our analysis, we will
need the sum coefficients $a_p$ for the modes $l=m=0$ and $l=m=1$. We start
with the case with $l=m=0$: employing again Eq. (\ref{eigen-sc}) and 
(\ref{coefficients}), we find the following results for the angular eigenvalue
\beq
_{0}A_{00}=- \frac{(a \omega)^2}{3}\,,
\eeq
and the $(\alpha_p, \beta_p, \gamma_p)$ coefficients
\beq
\alpha_p=-2 (p+1)^2\,, \qquad \beta_p=p(p+1) -a\omega(4p+2)-\frac{2(a\omega)^2}{3}\,,
\qquad \gamma_p=2a\omega p\,,
\eeq
that, in turn, lead to the following relations between the first five
sum coefficients $a_p$
\beq
\frac{a_1^{(00)}}{a_0^{(00)}}=-a \omega\,(1+\frac{a\omega}{3})\,,
\qquad
\frac{a_2^{(00)}}{a_0^{(00)}}=\frac{(a\omega)^2}{3}\left(2+ a\omega + 
\frac{(a\omega)^2}{12}\right)\,,
\eeq
\smallskip
\beq
\frac{a_3^{(00)}}{a_0^{(00)}}=-\frac{(a\omega)^3}{3}\left(1+
\frac{65 a\omega}{108}\right)+...\,, \qquad
\frac{a_4^{(00)}}{a_0^{(00)}}=\frac{(a\omega)^4}{108}\left(\frac{115}{8}+
\frac{29 a\omega}{3}\right)+...\,.
\label{a_000}
\eeq

\noindent
For the mode $l=m=1$, a similar analysis leads to the following results for
the eigenvalue
\beq
_{0}A_{11}=2- \frac{(a \omega)^2}{5}\,,
\eeq
the $(\alpha_p, \beta_p, \gamma_p)$ coefficients
\beq
\alpha_p=-2 (p+1)(p+2)\,, \qquad \beta_p=p(p+3) -4a\omega(p+1)-\frac{4(a\omega)^2}{5}\,,
\qquad \gamma_p=2a\omega (p+1)\,,
\eeq
and the first five sum coefficients $a_p$
\beq
\frac{a_1^{(11)}}{a_0^{(11)}}=-a \omega\,(1+\frac{a\omega}{5})\,,
\qquad
\frac{a_2^{(11)}}{a_0^{(11)}}=\frac{(a\omega)^2}{5}\left(3+ a\omega + 
\frac{(a\omega)^2}{15}\right)\,,
\eeq
\smallskip
\beq
\frac{a_3^{(11)}}{a_0^{(11)}}=-\frac{(a\omega)^3}{15}\left(4+
\frac{103 a\omega}{60}\right)+...\,, \qquad 
\frac{a_4^{(11)}}{a_0^{(11)}}=\frac{(a\omega)^4}{300}\left(\frac{571}{20}+
\frac{43 a\omega}{3}\right)+...\,.
\label{a_011}
\eeq


\begin{thebibliography}{99}

\bibitem{ADD}
N.~Arkani-Hamed, S.~Dimopoulos and G.~R.~Dvali,
{\it Phys.\ Lett.}\ B {\bf 429}, 263 (1998) [hep-ph/9803315];
{\it Phys.\ Rev.}\ D {\bf 59}, 086004 (1999) [hep-ph/9807344];
\\
I.~Antoniadis, N.~Arkani-Hamed, S.~Dimopoulos and G.~R.~Dvali,
{\it Phys.\ Lett.}\ B {\bf 436}, 257 (1998) [hep-ph/9804398].

\bibitem{BF} T.~Banks and W.~Fischler,
hep-th/9906038. 

\bibitem{hawking}
S.~W.~Hawking,
{\it Commun.\ Math.\ Phys.}\ {\bf 43}, 199 (1975).
 
\bibitem{Kanti}
P.~Kanti,
{\it Int. J. Mod. Phys.}\ A {\bf 19}, 4899 (2004) [hep-ph/0402168];
{\it {Lect.\ Notes Phys.\ }}  {\bf 769}, 387 (2009) [arXiv:0802.2218 [hep-th]];
 {\it J.\ Phys.\ Conf.\ Ser. }  {\bf 189} (2009) 012020  [arXiv:0903.2147 [hep-th]];
{\it Rom.\ J.\ Phys.}  {\bf 57} (2012) 96  [arXiv:1204.2371 [hep-th]].

\bibitem{reviews}
M.~Cavaglia,
  {\it {Int.\ J.\ Mod.\ Phys.}}\ A {\bf 18}, 1843 (2003)
  [hep-ph/0210296];
  \\
  G.~L.~Landsberg,
  {\it {Eur.\ Phys.\ J.}}\ C {\bf 33}, S927 (2004)
  [hep-ex/0310034];
\\
K.~Cheung,
  hep-ph/0409028; 
\\
S.~Hossenfelder,
  hep-ph/0412265; 
\\
A.~S.~Majumdar and N.~Mukherjee,
  {\it {Int.\ J.\ Mod.\ Phys.}}\ D {\bf 14}, 1095 (2005)
  [astro-ph/0503473]; \\
E.~Winstanley,
  arXiv:0708.2656 [hep-th];  
\\
  S.~C.~Park,
  {\it Prog.\ Part.\ Nucl.\ Phys.}  {\bf 67} (2012) 617  [arXiv:1203.4683 [hep-ph]].
  

\bibitem{Harris}
C.~M.~Harris, hep-ph/0502005. 

\bibitem{kmr1}
P.~Kanti and J.~March-Russell,
{\it Phys.\ Rev.}\ D {\bf 66}, 024023 (2002) [hep-ph/0203223];
{\it Phys.\ Rev.}\ D {\bf 67}, 104019 (2003) [hep-ph/0212199].

\bibitem{Frolov1}
V.~P.~Frolov and D.~Stojkovic,
{\it Phys.\ Rev.}\ D {\bf 66}, 084002 (2002) [hep-th/0206046].

\bibitem{HK1}
C.~M.~Harris and P.~Kanti,
{\it JHEP} {\bf 0310}, 014 (2003) [hep-ph/0309054].

\bibitem{graviton-schw}
A.~S.~Cornell, W.~Naylor and M.~Sasaki,
{\it JHEP} {\bf {0602}}, 012 (2006) [hep-th/0510009]; 
\\
V.~Cardoso, M.~Cavaglia and L.~Gualtieri,
{\it Phys.\ Rev.\ Lett.}\ {\bf 96}, 071301 (2006) [hep-th/0512002];
{\it JHEP} {\bf 0602}, 021 (2006)  [hep-th/0512116];
\\
S.~Creek, O.~Efthimiou, P.~Kanti and K.~Tamvakis,
{\it Phys.\ Lett.}\ B {\bf 635}, 39 (2006) [hep-th/0601126];
\\
D.~C.~Dai, N.~Kaloper, G.~D.~Starkman and D.~Stojkovic,
  {\it {Phys.\ Rev.}}\  D {\bf 75}, 024043 (2007)
  [hep-th/0611184]. 

\bibitem{KGB}
P.~Kanti, J.~Grain and A.~Barrau,
{\it Phys.\ Rev.}\ D {\bf 71} (2005) 104002  [hep-th/0501148].  

\bibitem{GBK}
  J.~Grain, A.~Barrau and P.~Kanti,
{\it Phys.\ Rev.}\ D {\bf 72} (2005) 104016  [hep-th/0509128].  

\bibitem{Nicolini}
  P.~Nicolini and E.~Winstanley,
  JHEP {\bf 1111} (2011) 075
  [arXiv:1108.4419 [hep-ph]]. 

\bibitem{DHKW}
C.~M.~Harris and P.~Kanti,
{\it Phys.\ Lett.}\ B {\bf 633} (2006) 106  [hep-th/0503010];\\  
G.~Duffy, C.~Harris, P.~Kanti and E.~Winstanley,
{\it JHEP} {\bf {0509}}, 049 (2005) [hep-th/0507274].

\bibitem{CKW}
M.~Casals, P.~Kanti and E.~Winstanley,
{\it JHEP} {\bf 0602}, 051 (2006) [hep-th/0511163];

\bibitem{CDKW1}
M.~Casals, S.~Dolan, P.~Kanti and E.~Winstanley,
{\it JHEP} {\bf 0703}, 019 (2007) [hep-th/0608193].

\bibitem{IOP}
D.~Ida, K.~y.~Oda and S.~C.~Park,
{\it {Phys.\ Rev.}}\ D {\bf 67}, 064025 (2003)
[Erratum-ibid.\ D {\bf 69}, 049901 (2004)] [hep-th/0212108];
{\it Phys.\ Rev.}\  D {\bf 71}, 124039 (2005) [hep-th/0503052];
{\it Phys.\ Rev.}\ D {\bf 73}, 124022 (2006) [hep-th/0602188].

\bibitem{CEKT2}
  S.~Creek, O.~Efthimiou, P.~Kanti and K.~Tamvakis,
{\it Phys.\ Rev.}\ D {\bf 75} (2007) 084043  [hep-th/0701288].  

\bibitem{CEKT3}
 S.~Creek, O.~Efthimiou, P.~Kanti and K.~Tamvakis,
{\it Phys.\ Rev.}\ D {\bf 76} (2007) 104013  [arXiv:0707.1768 [hep-th]].  

\bibitem{rot-other}
V.~P.~Frolov and D.~Stojkovic,
{\it Phys.\ Rev.}\ D {\bf 67}, 084004 (2003) [gr-qc/0211055];
\\
H.~Nomura, S.~Yoshida, M.~Tanabe and K.~i.~Maeda,
{\it  Prog.\ Theor.\ Phys.\ }  {\bf 114}, 707 (2005)
  [hep-th/0502179].
\\
T.~Kobayashi, M.~Nozawa, Y.~Takamizu,
{\it {Phys.\ Rev.\ }} D {\bf 77}, 044022 (2008)
[arXiv:0711.1395 [hep-th]];
\\
S.~Chen, B.~Wang, R.~K.~Su and W.~Y.~Hwang,
  {\it {JHEP}} {\bf 0803}, 019 (2008)
  [arXiv:0711.3599 [hep-th]].

\bibitem{CEKT4}
S.~Creek, O.~Efthimiou, P.~Kanti and K.~Tamvakis,
{\it Phys. Lett.}\ B {{\bf 656}}, 102 (2007) [arXiv: 0709.0241 [hep-th]];

\bibitem{CDKW2}
M.~Casals, S.~R.~Dolan, P.~Kanti and E.~Winstanley,
  {\it {JHEP}} {\bf 0806}, 071 (2008)
  [arXiv:0801.4910 [hep-th]]. 

\bibitem{graviton-rot}
H.~Kodama,
  {\it {Prog.\ Theor.\ Phys.\ Suppl.\ }}  {\bf 172}, 11 (2008)
  [arXiv:0711.4184 [hep-th]];
  {\it {Lect.\ Notes Phys.}}\  {\bf 769}, 427 (2009)
  [arXiv:0712.2703 [hep-th]];
  \\
  J.~Doukas, H.~T.~Cho, A.~S.~Cornell and W.~Naylor,
  {\it Phys.\ Rev.}\ D {\bf 80} (2009) 045021  [arXiv:0906.1515 [hep-th]];  
\\
P.~Kanti, H.~Kodama, R.~A.~Konoplya, N.~Pappas and A.~Zhidenko,
{\it Phys.\ Rev.}\ D {\bf 80} (2009) 084016  [arXiv:0906.3845 [hep-th]].  

  
\bibitem{brane-bulk}
E.~Jung and D.~K.~Park,
{\it Nucl.\ Phys.}\ B {\bf{731}}, 171 (2005) [hep-th/0506204];
{\it Mod.\ Phys.\ Lett.}\ A {\bf{22}}, 1635 (2007) [hep-th/0612043].

\bibitem{Sampaio}
  M.~O.~P.~Sampaio,
{\it JHEP} {\bf 0910} (2009) 008  [arXiv:0907.5107 [hep-th]]; 
  {\it JHEP} {\bf 1002} (2010) 042  [arXiv:0911.0688 [hep-th]].  

\bibitem{KP1} P.~Kanti and N.~Pappas,
{\it Phys.\ Rev.}\ D {\bf 82} (2010) 024039  [arXiv:1003.5125 [hep-th]].

\bibitem{charybdis2}
J.~A.~Frost, J.~R.~Gaunt, M.~O.~P.~Sampaio, M.~Casals, S.~R.~Dolan,
M.~A.~Parker and B.~R.~Webber,
  JHEP {\bf 0910} (2009) 014  [arXiv:0904.0979 [hep-ph]].  

\bibitem{blackmax}
D-C.~Dai, G.~Starkman, D.~Stojkovic, C.~Issever, E.~Rizvi,
and J.~Tseng,
{\it {Phys.\ Rev.}}\  D {\bf 77}, 076007 (2008)
  [arXiv:0711.3012 [hep-ph]].

\bibitem{FST} A.~Flachi, M.~Sasaki and T.~Tanaka,
{\it JHEP} {\bf 0905} (2009) 031  [arXiv:0809.1006 [hep-ph]].  

\bibitem{CDKW3} 
M.~Casals, S.~R.~Dolan, P.~Kanti and E.~Winstanley,
{\it Phys.\ Lett.}\ B {\bf 680} (2009) 365  [arXiv:0907.1511 [hep-th]].

\bibitem{Sampaio-ang}
  M.~O.~P.~Sampaio,
  {\it JHEP} {\bf 1203} (2012) 066  [arXiv:1201.2422 [hep-ph]].  

\bibitem{Stojkovic-ang} D.~C.~Dai and D.~Stojkovic,
{\it JHEP} {\bf 1008} (2010) 016  [arXiv:1008.4586 [gr-qc]].  

\bibitem{MP}
R.~C.~Myers and M.~J.~Perry,
{\it Annals Phys.}\  {\bf 172}, 304 (1986).

\bibitem{Teukolsky}
S.~A.~Teukolsky,
 {\it {Phys.\ Rev.\ Lett.}}\ {\bf 29}, 1114 (1972);
  {\it {Astrophys.\ J.}}\ {\bf 185}, 635 (1973).

  
\bibitem{press1} W.~H.~Press and S.~A.~Teukolsky,
Astrophys.\ J.\  {\bf 185}, 649 (1973). 

\bibitem{fackerell}
E.~D.~Fackerell and R.~G.~Grossman,
J.\ Math.\ Phys.\ {\bf 18}, 1849 (1977).

\bibitem{churilov} A.~A.~Starobinskii and S.~M.~Churilov,
{\it Sov. Phys.-JETP} {\bf 38}, 1 (1974).

\bibitem{Seidel} E.~Seidel,
Class.\ Quant.\ Grav.\  {\bf 6}, 1057 (1989).

\bibitem{BCC}
E.~Berti, V.~Cardoso and M.~Casals,
Phys.\ Rev.\  D {\bf 73} (2006) 024013 [Erratum-ibid.\  D {\bf 73}
(2006) 109902];

\bibitem{Leaver} 
E.~W.~Leaver, Proc.\ Roy.\ Soc.\ London A {\bf {402}}, 285 (1985).


\end{thebibliography}
\end{document}